\documentclass[12pt]{article}
\usepackage{latexsym,epsfig,graphicx,epstopdf,amsmath,amssymb,amscd,multirow,psfrag,paralist,dsfont,url,setspace,bm,algpseudocode,pifont,caption,subcaption,booktabs,etoolbox}
\usepackage{chicago}

\usepackage[titletoc]{appendix}
\usepackage{booktabs}
\usepackage{epstopdf}
\usepackage[american]{babel}
\usepackage{xr}

\renewcommand{\arraystretch}{.8}

\newcommand{\indi}{\boldsymbol{\mathds{1}}}

\newcommand{\normal}{\mathcal{N}}

\newcommand{\IG}{\mathcal{IG}}
\newcommand{\E}{\bold{E}}
\newcommand{\trunnor}{\mathcal{N}_{\text{trunc}}}
\newcommand{\likh}{\mathcal{L}}
\newcommand{\loglikh}{\ell}

\newcommand{\Phimat}{\Phi}
\newcommand{\Sigmat}{\Sigma}
\newcommand{\Omegamat}{\Omega}

\newcommand{\makematrixcmd}[1]{%
	\expandafter\newcommand\csname #1mat\endcsname{\mathbf{#1}}%
}

\forcsvlist{\makematrixcmd}{A,B,C,D,E,F,G,H,I,J,K,L,M,N,O,P,Q,R,S,T,U,V,W,X,Y,Z}
\newcommand{\tran}{^\top}
\newcommand{\diag}{\textrm{Diag}}

\newcommand{\alphavec}{\boldsymbol{\alpha}}
\newcommand{\gammavec}{\boldsymbol{\gamma}}

\newcommand{\betavec}{\boldsymbol{\beta}}
\newcommand{\thetavec}{\boldsymbol{\theta}}

\newcommand{\etavec}{\boldsymbol{\eta}}

\newcommand{\muvec}{\boldsymbol{\mu}}
\newcommand{\kappavec}{\boldsymbol{\kappa}}
\newcommand{\upsilonvec}{\boldsymbol{\upsilon}}
\newcommand{\zetavec}{\boldsymbol{\zeta}}

\newcommand{\zerovec}{{\boldsymbol{0}}}
\newcommand{\onevec}{{\boldsymbol{1}}}

\newcommand{\makevectorcmd}[1]{%
	\expandafter\newcommand\csname #1vec\endcsname{\boldsymbol{#1}}%
}
\forcsvlist{\makevectorcmd}{A,B,C,D,E,F,G,H,I,J,K,L,M,N,O,P,Q,R,S,T,U,V,W,X,Y,Z}

\newcommand{\makeveccmd}[1]{%
	\expandafter\newcommand\csname #1vec\endcsname{\boldsymbol{#1}}%
}

\forcsvlist{\makeveccmd}{a,b,c,d,e,f,g,h,i,j,k,l,m,n,o,p,q,r,s,t,u,v,w,x,y,z}

\newcommand{\makesetcmd}[1]{%
	\expandafter\newcommand\csname #1set\endcsname{\mathcal{#1}}%
}
\forcsvlist{\makesetcmd}{A,B,C,D,E,F,G,H,I,J,K,L,M,N,O,P,Q,R,S,T,U,V,W,X,Y,Z}

\newtheorem{preprop}{Proposition}
{\begin{preprop}\upshape}{\end{preprop}}

\newcommand{\Dfun}{\mathcal{D}}
\newcommand{\gfun}{g}
\newcommand{\ffun}{f}

\newcommand\restr[2]{{
		\left.\kern-\nulldelimiterspace 
		#1 
		\vphantom{\big|} 
		\right|_{#2} 
}}

\renewcommand{\arraystretch}{.8}

\begin{document}

\title{Modeling Multivariate Degradation Data with Dynamic Covariates Under a Bayesian Framework}

\author{
Zhengzhi Lin$^{1}$, Xiao Liu$^{2}$, Yisha Xiang$^{3}$, and Yili Hong$^{1}$\\[1.5ex]
{\small $^1$Department of Statistics, Virginia Tech, Blacksburg, VA 24060, USA}\\
{\small $^{2}$H. Milton Stewart School of ISE, Georgia Tech, Atlanta, GA 30332, USA}\\
{\small $^{3}$Department of Industrial Engineering, University of Houston, Houston, TX 77004, USA}
}

\date{}

\maketitle

\begin{abstract}
Degradation data are essential for determining the reliability of high-end products and systems, especially when covering multiple degradation characteristics (DCs). Modern degradation studies not only measure these characteristics but also record dynamic system usage and environmental factors, such as temperature, humidity, and ultraviolet exposures, referred to as the dynamic covariates. Most current research either focuses on a single DC with dynamic covariates or multiple DCs with fixed covariates. This paper presents a Bayesian framework to analyze data with multiple DCs, which incorporates dynamic covariates. We develop a Bayesian framework for mixed effect nonlinear general path models to describe the degradation path and use Bayesian shape-constrained P-splines to model the effects of dynamic covariates. We also detail algorithms for estimating the failure time distribution induced by our degradation model, validate the developed methods through simulation, and illustrate their use in predicting the lifespan of organic coatings in dynamic environments.


\end{abstract}

\newpage

\section{Introduction}
\label{sec:intro}

\subsection{Background}
\label{sec:background}
Degradation data are widely used in reliability analysis of products, systems, and materials. One advantage of degradation data, as compared to time to event data, is that it can be used to assess reliability of a product before it fails. The modeling and analysis of degradation data enable researchers to predict the lifespan of a product, components, and materials, and determine the factors that contribute to its failure. Two main factors, usage and environmental conditions, contribute to the degradation of a product. Environmental conditions refer to various factors that impact the degradation path of a product. These factors typically include covariates such as temperature, humidity, and other relevant information. Accounting for these environmental conditions can often be challenging for reliability analysis. Nonetheless, incorporating the environmental conditions in analysis is crucial for predicting the lifespan of a product and ensuring its reliability. Therefore, a comprehensive understanding of the impact of environmental conditions on the degradation path is vital for reliability analysis (e.g., Chapter~1 of \citeNP{Meeker2022}).

In laboratory experiments, we can manage specific environmental factors. In field studies, however, controlling these factors is challenging, especially when they change over time, which are known as dynamic covariates. There are two main reasons to include dynamic covariates in modeling degradation paths. With advancements in technology, collecting accurate covariate information and degradation measurements from field products has become more cost-effective. Additionally, dynamic covariates provide a richer amount of information, which could lead to a more accurate and comprehensive understanding of the degradation paths.

A product may exhibit multiple degradation characteristics (DCs) that capture our interest. Each DC corresponds to a distinct degradation path, but all share common environmental information. The degradation paths within the same product unit is likely to be correlated. Therefore, modeling multivariate degradation paths is essential since it helps capture the correlations between multiple DCs.

In this paper, we propose a method to model the degradation paths of multivariate degradation data (i.e., degradation data with multiple DCs) with dynamic covariates. Our method utilizes a Bayesian framework to incorporate information from multiple sources, including the degradation data and the dynamic covariates, to carry out reliability analysis. We implement Bayesian P-splines into the framework and propose Bayesian shape constraints to ensure that the covariate effect functions align with the expected shape assumptions. We validate our proposed method using simulation studies. To demonstrate our method's effectiveness in modeling the degradation path of products under dynamic covariates, we apply it to a real-world case study. Our findings highlight the importance of considering dynamic covariates in reliability analysis and the developed framework provides useful tools for practitioners in this field.

\subsection{Motivating Application: Outdoor Weathering Study} \label{sec:motivating.application}

\begin{figure}
	\centering
	\begin{subfigure}{0.3\textwidth}
		\centering
		\includegraphics[width=\linewidth]{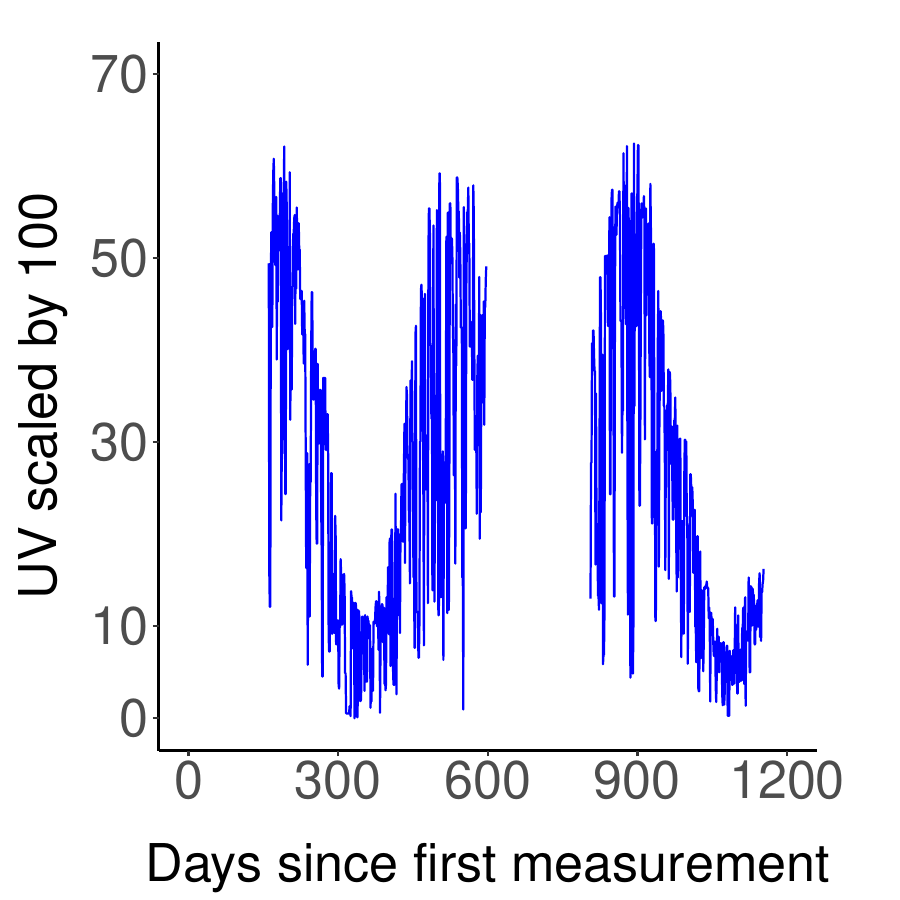}
		\caption{Daily UV dosage}
	\end{subfigure}%
	\begin{subfigure}{0.3\textwidth}
		\centering
		\includegraphics[width=\linewidth]{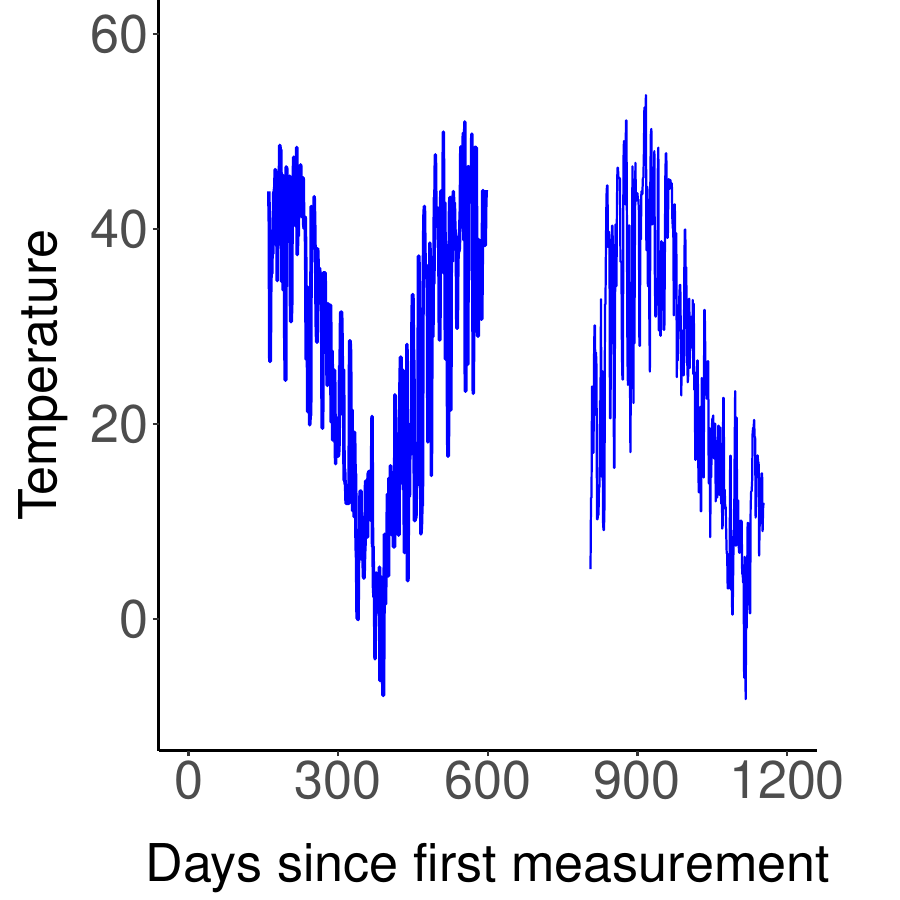}
		\caption{Daily TEMP}
	\end{subfigure}%
	\begin{subfigure}{0.3\textwidth}
		\centering
		\includegraphics[width=\linewidth]{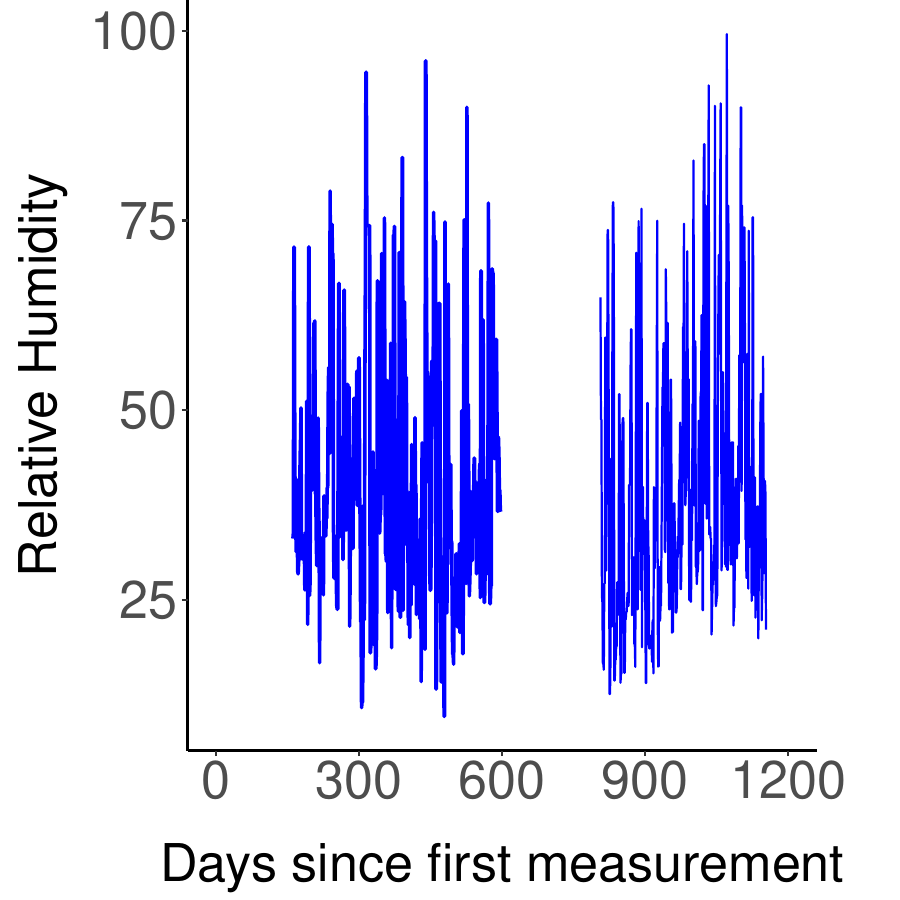}
		\caption{Daily RH}
	\end{subfigure}
	\caption{Plots for three dynamic covariates UV, TEMP, and RH from 01 January 2002 to 23 August 2003 and from 28 September 2004 to 04 July 2005. The gaps in the plots are due to a pause during the study. }
	\label{fig:cov.process}
\end{figure}

The motivation of this paper is from an outdoor weathering dataset, derived from a field study conducted by the National Institute of Standards and Technology (NIST) (\shortciteNP{Guetal2009}), which focused on gathering essential lifetime data for organic coatings utilized in outdoor utilities. We refer to the dataset as the weathering data for the rest of this paper.  The study, conducted in Gaithersburg, MD from 2002 to 2006, placed 36 distinct specimens in an outdoor chamber to examine environmental conditions. The specimens were situated on the NIST campus, with each starting at varying times across a five-year duration. Throughout this period, dynamic environmental covariates, such as temperature (TEMP), humidity (RH), and ultraviolet (UV) intensity, were monitored and recorded automatically using sensors. Figure~\ref{fig:cov.process} exhibits the covariate data collected over the period of the study, highlighting the seasonal patterns of UV and TEMP variables. In contrast, the RH variable presents a less seasonal pattern and a considerable higher degree of variance. The majority of data points for RH lie within the 25 to 50 range, whereas fewer points are observed above 50 or below 25. The presence of such substantial variance in the covariate information may lead to challenges in the modeling process, which will be discussed later.

In the coating study, the researchers periodically measured the degradation of coating specimens by employing Fourier transform infrared (FTIR) spectroscopy. Chemical compounds in the coating absorb light at specific wavelengths, producing distinct peaks in the FTIR spectrum, with peak height corresponding to the compound's concentration. As degradation occurs, the concentration decreases, resulting in alterations to the peak height on the FTIR spectrum, which can thus serve as an indicator for degradation (i.e., damage). For each specimen involved in this study, three distinct degradation paths are measured using different chemical molecule stretching at wavelength 1250 $\mbox{cm}^{-1}$, at 1510 $\mbox{cm}^{-1}$, and at 2925 $\mbox{cm}^{-1}$. We denote the three DCs by their corresponding wavelength for convenience.

\begin{figure}
	\centering
	\begin{subfigure}{0.33\textwidth}
		\centering
		\includegraphics[width=\linewidth]{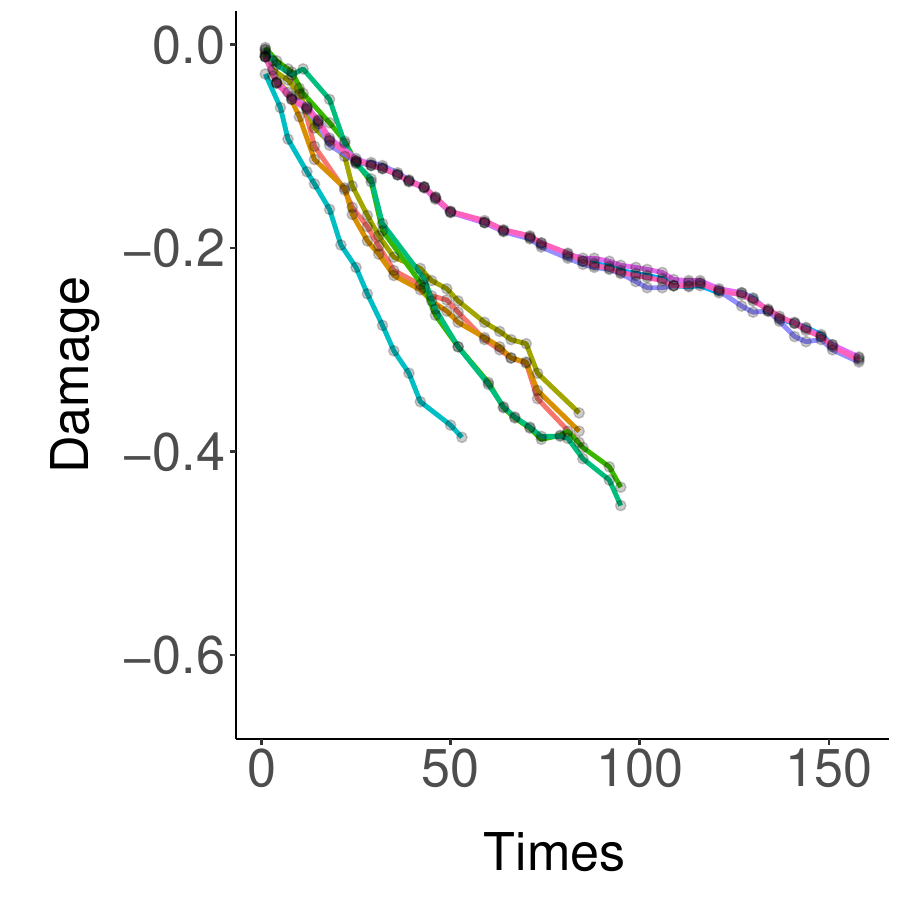}
		\caption{1250 $\mbox{cm}^{-1}$}
	\end{subfigure}%
	\begin{subfigure}{0.33\textwidth}
		\centering
		\includegraphics[width=\linewidth]{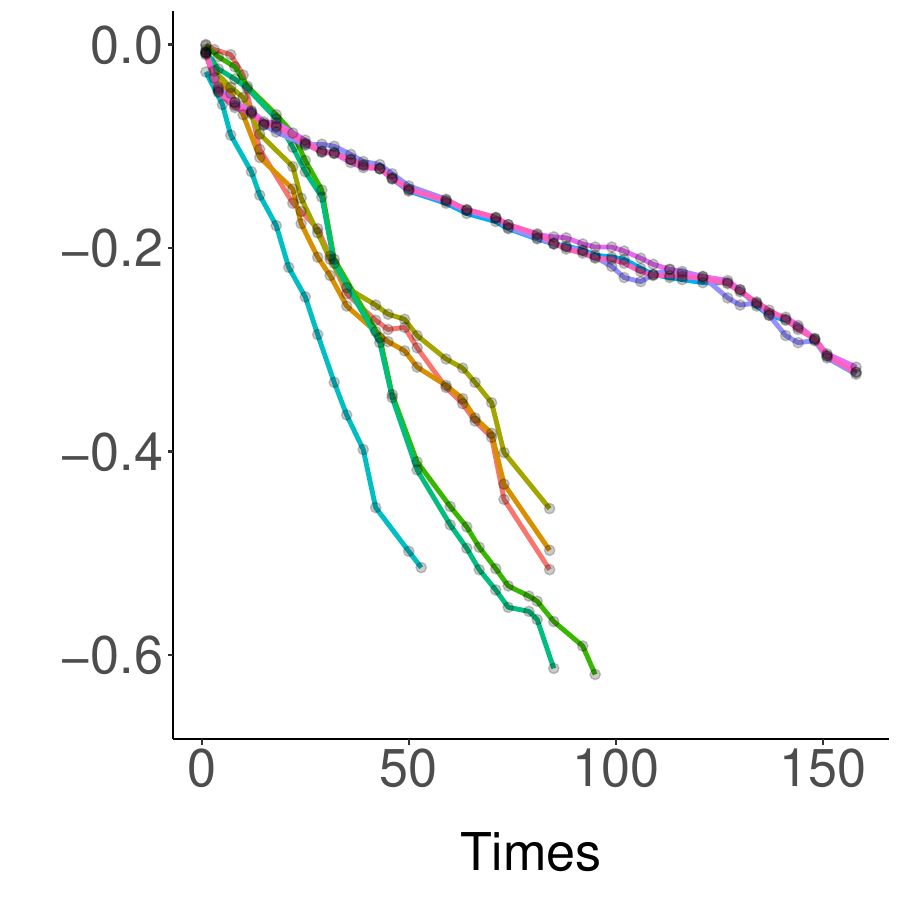}
		\caption{1510 $\mbox{cm}^{-1}$}
	\end{subfigure}%
	\begin{subfigure}{0.33\textwidth}
		\centering
		\includegraphics[width=\linewidth]{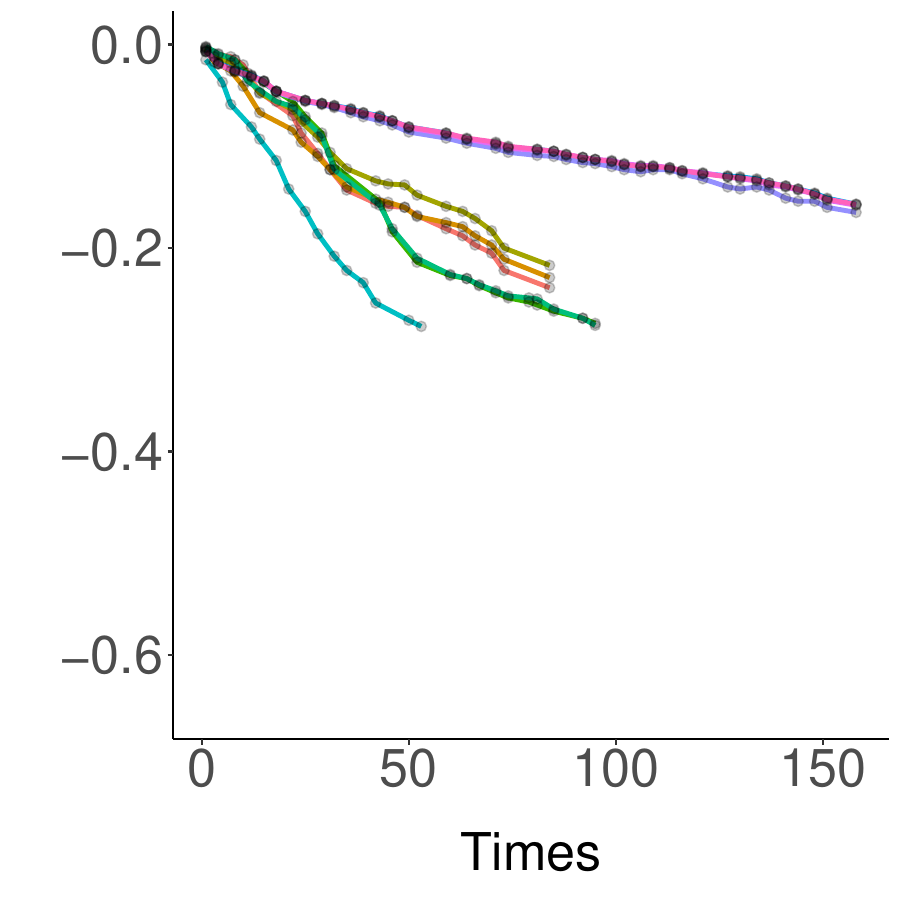}
		\caption{2925 $\mbox{cm}^{-1}$}
	\end{subfigure}
	\caption{	Coating degradation paths for ten representative specimens. Each color corresponds to a different specimen. The $x$-axis represents the time in days since the first measurement.}
	\label{fig:deg.damage}
\end{figure}

Figure \ref{fig:deg.damage} displays the degradation paths for 10 selected specimens out of 36, with each path being measured under different wavelengths. Each specimen has a unique starting date, and the $x$-axis represents the number of days elapsed since the initial measurement of the respective specimen. Due to different starting times, each specimen owns distinct covariate profiles, resulting in different rates of degradation as shown in Figure \ref{fig:deg.damage}.

\subsection{Related Literature and Contributions of This Work}
\label{sec:related.literature}
The general path models are popular in degradation modeling due to their ability to capture sources of variations in a degradation process. The general path models are a class of nonlinear mixed-effects models that are designed to model nonlinear degradation paths and their functional forms are usually suggested by domain knowledge from the field. \citeN{Meeker1993} provided a detailed definition of the general path model and derived the cumulative lifetime distribution.  In degradation data analysis, various settings have been developed to incorporate covariate information into modeling and subsequent inference. These include accelerated repeated measure degradation tests (e.g., \citeNP{Nikulin2001}), accelerated destructive degradation tests (e.g., \shortciteNP{Escobar2003}), and experimental designs (e.g., \citeNP{Padgett2006}). In addition to general path models, there have been numerous developments in stochastic models for modeling degradation data. For example, \shortciteN{WANG2020106631} considered reliability inference using a Wiener process with random effects. \shortciteN{LI2023109078}  integrated machine learning and stochastic models for degradation modeling. \shortciteN{LI2024110101} developed accelerated degradation testing for lifetime analysis with random effects.

To model dynamic covariates, stochastic degradation models were proposed (\shortciteNP{Nikulin2010}).  \shortciteN{Hong2015} introduced a mixed linear model that integrates dynamic covariates with random effects to account for variations between specimens. The covariate information is described through shape-constrained regression splines introduced by \citeN{Meyer2008}. \citeN{Xu2016} proposed a nonlinear general path model with a similar covariate modeling scheme. In both work, the selecting of spline knots can be challenging and the rigidity of shape constraints can lead to potential problems when applying to other data. Several researchers have offered solutions to address these issues. \citeN{Marx1996} introduced P-splines, eliminating the need for knot selection; however, they require validation for penalty parameter choice. \citeN{Lang2004} suggested the use of Bayesian P-splines, which resolves the penalty parameter selection dilemma by assigning a non-informative inverse gamma prior to the parameter. To obtain the desired shape with the original P-splines, \citeN{Bollaerts2006} suggested adding a multi-degree penalty term to penalize segments with undesired shapes. In contrast, \citeN{Pya2015} introduced a shape-restricted additive model that achieves shapes by implementing specific linear transformations on spline coefficients. Building on Bayesian P-splines, \citeN{Brezger2012} presented a variation of Bayesian P-splines designed to ensure monotonicity.

With the rise of degradation data with multiple DCs, there has been a growing call for models that can accommodate such features. To model the correlation among DCs, \shortciteN{Pan2013} and \shortciteN{Wang2015}  used copula-based approaches.  \citeN{FANG2020106618} developed copula-based reliability inference and \citeN{FANG20221177} developed inverse Gaussian processes for multiple DCs. \shortciteN{Lu2021} introduced a model that can handle multiple DCs with fixed covariates. \citeN{li2023system} introduced several degradation models and modeling approaches for reliability analysis to study dependent processes. \shortciteN{10287198} considered multi-dimensional degradation processes for modeling degradation of high-speed train wheels.
\citeN{FangPan2024} developed multivariate Wiener processes for modeling dependent degradation data. \shortciteN{ASGARI2024110146} proposed to use multivariate fractional Brownian motion to model multivariate degradation data.

In practice, the simultaneous presence of multiple DCs and dynamic covariates is a more common scenario, which has not been addressed in existing work. To enhance degradation prediction and understand degradation mechanisms influenced by dynamic covariates, there's a need for a model that accommodates both aspects. In this paper, we introduce a model that captures the complex nonlinear nature of degradation process, integrating multiple DCs, dynamic covariates, and specimen-specific random effects. Our model extends the nonlinear general path models from \shortciteN{Lu2021} and \citeN{Xu2016}, utilizing shape-constrained Bayesian P-splines within the Bayesian framework to manage overfitting and adhere to shape constraints. Our spline configurations are adaptable and can be applied to similar applications. Results from our simulation study confirm the efficacy of our method in modeling dynamic covariates with both monotonic and convex shapes.

\subsection{Overview}
\label{sec:overview}
The remainder of this paper is structured as follows. Section~\ref{sec:data.models} introduces our notations for degradation data, the modeling of the covariate effects, and the general path model. Section~\ref{sec:model.est} proposes the Bayesian framework for our model, introduces the shape-constrained Bayesian P-splines for modeling the covariate effects, describes the way we conduct posterior inference, and introduces an algorithm for estimating failure time distributions. Section~\ref{sec:simulations} shows our simulation study that validates the model performance compared to the independent degradation model. Section~\ref{sec:application} demonstrates the application of our method on the weathering data. Section~\ref{sec:conclusions} lists our conclusions and directions for future research.

\section{Data and Statistical Models}\label{sec:data.models}
\subsection{Data and Notations}
Let $i=1,\dots, I$ be the index for those $I$ specimens involved in the study and let $j=1,\dots,J$ be the index for those $J$ different types of DCs that are under measurement. For a given specimen, the time points for measuring its degradation could differ from those for recording its associated covariate information. To represent this, let $\Tset_i = \left\{t_{ik}, k=1,\dots,n_i\right\}$ denote the set of time points for measuring the DCs of specimen $i$, where $n_i$ is the number of degradation measurement points. Similarly, let $\Sset_i=\left\{s_{il}, l=1,\dots, l_i\right\}$ denote the time points at which covariate information for the specimen is recorded, and $l_i$ is the number of covariate measuring points. Let $y_{ijk}$ be the degradation measurement for the $j$th DC of specimen $i$ at time $t_{ik}$.

Suppose there are $M$ different covariates, which are indexed by $m$. Denote the covariate value for specimen $i$, covariate $m$, at time $s \in \Sset_i$ as $x_{im}(s)$. Denote covariate process history up to time $t$ as $\xvec_{im}(t) = \left\{x_{im}(s): 0 \leq s \leq t, s\in \Sset_i \right\}$. Then all covariate's history of specimen $i$ can be denoted as $\Xmat_{i}(t) = \left\{\xvec_{im}(t), m=1,\dots,M \right\}$. The covariate process for all specimens up to time $t$ is denoted as $\Xmat(t)=\left\{\Xmat_i(t), i=1,\dots,I\right\}$.

\subsection{General Path Model}
The general degradation path model for degradation measurement $y_{ijk}$ is
\begin{align}\label{eqn:gpm.genral.form}
	y_{ijk} = \Dfun_{j}(t_{ik}) + \epsilon_{ijk}, \quad i=1,\dots,I, \quad j=1,\dots,J, \quad k= 1,\dots, n_i,
\end{align}
where $\epsilon_{ijk} \sim \normal(0, \sigma^2)$ is an error term with variance $\sigma^2$. The function $\Dfun_j(t_{ik})$ represents the true degradation path for the $j$th DC at time $t_{ik}$, which is influenced by the dynamic covariate history, including all historical data for the dynamic covariate process up to time $t_{ik}$. Without loss of generality, we assume $\Dfun_{j}(t)$ is a decreasing function of $t$.
In the following, we will develop a functional form for the covariate history and detail the modeling of $\Dfun_j(t_{ik})$.

We will build a connection between $\Dfun_j(t_{ik})$ and $\Xmat(t_{ik})$ by introducing the covariate effect function, the instant effect function, and the cumulative effect function. The covariate effect function is defined as a non-negative function that determines the amount of exposure a specimen receives under a known covariate value. The instant effect function is defined as the exposure a specimen receives for a short period of time. While the cumulative effect function is defined as the total amount of exposure a specimen receives for a duration of time.
The degradation path of a specimen up to time $t$ is a function of the cumulative exposure incurred from continuous exposure to covariates until time $t$. This is because exposing to environmental covariates is the main cause for degradation. We will first introduce the shape-constrained Bayesian P-splines that approximate the covariate effect function and satisfy shape assumptions from domain knowledge.

\subsection{Modeling Covariate Process}\label{subsec:covprocess}

\begin{figure}
	\centering
	\includegraphics[width=0.45\linewidth]{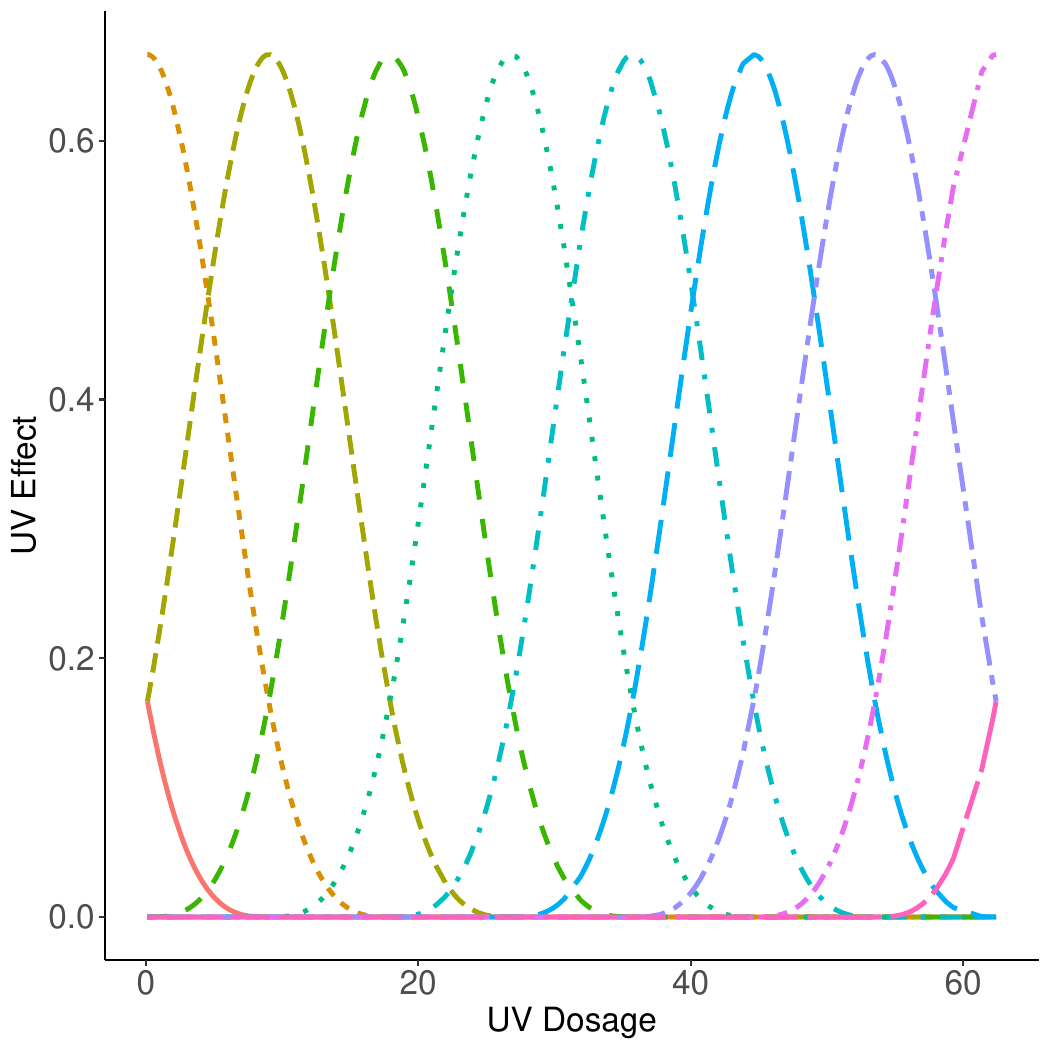}
	\caption{Illustration of B-spline basis functions for the UV covariate.}
	\label{fig:spline_basis.plt}
\end{figure}

The exposure that the $m$th covariate contributes to the $j$th DC is called the covariate effect, and it is denoted as $\ffun_{jm}(x)$, where $x$ is the covariate value. The function $\ffun_{jm}(x)$ is called the covariate effect function of the $m$th covariate on DC $j$, or the effect function in short. In this paper, we use a linear combination of B-splines to model the effect function. The B-spline basis functions are computed using equal-spaced knots (e.g., \citeNP{Marx2015}), which will be used later in P-splines. Figure~\ref{fig:spline_basis.plt} uses covariate UV as an example to illustrate the B-spline basis functions.

Here we introduce some notation. For the $j$th DC and covariate $m$, we use $Q_m$ number of non-negative B-spline basis functions $B_{mq}(x)$. The corresponding coefficients are $\beta_{jmq}$, $q=1,\dots,Q_m$, to form the approximation. In this paper, we select degree 3 for the B-spline basis functions to obtain the desired level of smoothness. Given the parameters and basis functions, the effect function of covariate $m$ is $\ffun_{jm}(x) = \sum_{q=1}^{Q_m} B_{mq}(x) \beta_{jmq}$.
Note that all $\beta_{jmq}$'s, $q=1,\dots, Q_m$, are positive due to the non-negativity of $\ffun_{jm}(x)$. In practice, an exponential transformation can be applied to enforce the positivity of $\beta_{jmq}$.
Specifically, suppose that specimen $i$ has covariate value $x_{im}(s)$ for covariate $m$ at time $s$, then
\begin{align*}
	\ffun_{jm}[x_{im}(s)] = \sum_{q=1}^{Q_m} B_{mq}[x_{im}(s)] \beta_{jmq}.
\end{align*}

The instant effect specimen $i$ receives due to covariate $m$ on the $j$th DC, or the instant effect function in short, can be expressed as $\ffun_{jm}[x_{im}(s)] ds$, which is the total amount of exposure specimen $i$ receives during the time interval $[s, s+ds)$. Denote $\gfun_{ijm}(t)$ as the cumulative effect function of covariate $m$ on specimen $i$ for DC $j$, or the cumulative effect function in short. It represents the cumulative exposure specimen $i$ receives due to covariate $m$ on the $j$th DC over the duration from time 0 to time $t$. Consequently, the cumulative effect function can be defined as an integral of the effect function as follows,
\begin{align*}
	\gfun_{ijm}(t) = \int_{0}^{t} \ffun_{jm}[x_{im}(s)] ds
	= \int_{0}^{t} \sum_{q=1}^{Q_m}B_{mq}[x_{im}(s)]\beta_{jmq} ds.
\end{align*}
In practice, the measuring time points are discrete. Thus for any $t_{ik} \in \Tset_i$, $g_{ijm}(t_{ik})$ can be approximated as a summation as follows,
\begin{align*}
	g_{ijm}(t_{ik}) 
	& = \sum_{s_{il} \leq t_{ik}} \sum_{q=1}^{Q_m}B_{mq}[x_{im}(s_{il})]\beta_{jmq}(s_{il} - s_{i,l-1}) 
	= \sum_{q=1}^{Q_m} G_{imq}(t_{ik}) \beta_{jmq},
\end{align*}
where $G_{imq}(t_{ik}) =\sum_{s_{il} \leq t_{ik}} B_{mq}[x_{im}(s_{il})](s_{il} - s_{i,l-1})$. Further we denote the $Q_m \times 1$ vector $\Gvec_{im}(t_{ik}) = \left(G_{im1}(t_{ik}), G_{im2}(t_{ik}), \dots, G_{imQ_{m}}(t_{ik})\right)\tran$,
and we call it the cumulative effect vector for specimen $i$ and covariate $m$ up to time $t_{ik}$. It captures the cumulative effect that a covariate has on a specimen over time. Therefore, $g_{ijm}(t_{ik})$ is the cumulative effect vector multiplies the coefficient vector $\betavec_{jm}=(\beta_{jm1}, \dots, \beta_{jmQ_{m}})\tran$. That is, $g_{ijm}(t_{ik}) = \Gvec_{im}(t_{ik})\tran\betavec_{jm}$.

\subsection{Modeling Degradation Path}\label{subsec:model}
Define the total cumulative effect
\begin{align*}
	\tau_{ijk} = \tau_{j}(t_{ik}) & = \sum_{m=1}^{M} g_{ijm}(t_{ik}) =  \sum_{m=1}^{M} \sum_{q=1}^{Q_m} G_{imq}(t_{ik})\beta_{jmq},
\end{align*}
as the sum of all cumulative exposure from covariates that specimen $i$ experiences on the $j$th DC up to time $t_{ik}$. With the above discussion, the model in \eqref{eqn:gpm.genral.form} can be written as,
\begin{align}\label{eqn:GPM.model.spline.re}
	y_{ijk} = \Dfun(\tau_{ijk};\thetavec_j, w_{ij}) + \epsilon_{ijk}.
\end{align}
Here, $\betavec_j = (\beta_{j11}, \dots, \beta_{j1Q_1}, \dots, \beta_{jM1}, \dots, \beta_{jMQ_M})\tran$ is the DC-specific covariate coefficient vector and
 $\thetavec_j$ is the vector of fixed parameters, which includes spline coefficients $\betavec_j$ and other model specified fixed parameters. Note that here we introduce random effect $w_{ij}$ to represent specimen variation and possible variation due to different DCs. Let $\wvec_{i} = (w_{i1}, \dots, w_{iJ})\tran$ denote the random effect vector. A specific functional form of $\Dfun(\cdot)$ is needed. For the weathering data, we modify the form of the degradation path in \citeN{Xu2016} by expanding it to the multiple DC scenario. Specifically, for the $j$th DC,
\begin{align*}
	\Dfun_{j}(t_{ik}) = \Dfun(\tau_{ijk};\thetavec_j, w_{ij}) &= -\frac{\alpha_j \exp(w_{ij})}{1+\exp\left( -\frac{\log(\tau_{ijk})}{\gamma_j}\right)}, \quad j=1,\dots, J,
\end{align*}
where $\thetavec_j = (\alpha_j, \gamma_j, \betavec_j\tran)\tran$, $\alpha_j$ and $\gamma_j$ are additional parameters in $\Dfun(\cdot)$. $\gamma_j$ determines the rate of degradation. The term $-\alpha_j \exp(w_{ij})$ gives the asymptote of the true degradation for the $j$th DC for specimen $i$ as $t$ goes to $\infty$.

The random effects vector $\wvec_{i}$ is assumed to have a multivariate normal distribution with mean $\zerovec_{J\times1}$ and covariance matrix $\Sigmat$. That is, $\wvec_{i}  \sim \normal\left(\zerovec, \Sigmat\right)$.
The $J \times J$ covariance matrix $\Sigmat$ can be decomposed as $\Sigmat = \Dmat \Omegamat \Dmat\tran$,
where $\Dmat = \diag(\sigma_{1}, \dots, \sigma_{J})$ and the correlation matrix $\Omegamat$ is
\begin{align*}
	\Omegamat = \begin{pmatrix}
		1 & \rho_{12} & \dots & \rho_{1J} \\
		\rho_{12} & 1 & \dots & \rho_{2J} \\
		\vdots & \vdots & \ddots & \vdots \\
		\rho_{1J} & \rho_{2J} & \dots & 1
	\end{pmatrix}.
\end{align*}
The element $\rho_{jj'}$, $j, j' = 1,\dots, J$, $j\neq j'$ introduces correlation between the $j$th DC and the $j'$th DC.

\section{P-Splines, Shape Constraint and Bayesian Inference}\label{sec:model.est}
In Section~\ref{sec:data.models}, we discussed a model for multiple DCs and specified the parameters by incorporating random effects. However, there are still several challenges that need to be addressed. First, we need to address the issue of overfitting, which is a common and critical concern when using splines. Second, implementing shape constraints is essential to ensure our estimated effect functions align with our physical knowledge. To address the first challenge, we propose to use Bayesian P-splines, which will be discussed in Section~\ref{subsec:bayes.pspline}. Based on the Bayesian P-splines framework, we then introduce our shape constraints to ensure the estimated effect function exhibits the desired shape, which will be discussed in Section~\ref{subsec:shape.constraint}. We will also discuss Bayesian inference and reliability estimation.

\subsection{The Bayesian P-Splines} \label{subsec:bayes.pspline}

Here we use a linear regression model to convey the idea of Bayesian P-splines. Let $x$  be the covariate. The initial step in the fitting process involves selecting the number of spline bases, denoted by $q$. Following this, the least squares estimation can be carried out by minimizing the term $(\yvec - \Xmat\betavec)\tran(\yvec - \Xmat\betavec)$, where $\yvec$ denotes the responses, $\Xmat$ is the spline matrix based on $x$, and $\betavec$ consists of the spline coefficients.

When implementing splines, selecting the appropriate $q$ can be challenging due to potential overfitting issues. \citeN{Marx1996} introduced P-splines to address the issue of overfitting in spline regression. Within the P-splines framework, an abundance number of B-spline basis functions are employed, accompanied by a penalty term designed to penalize the overly oscillatory components of the curve. The penalty term is a composite of a tuning parameter, $\lambda$, and the $L_2$ sum of a selected order of difference among the spline coefficients. Employing different orders of differences results in varying levels of smoothness in the fitted spline. Higher-order differences yield smoother curves, while lower-order differences allow for more flexibility in capturing local variations. The first and second orders of differences usually provide a sufficient level of smoothness for most applications.

Using regression as an example, with coefficients $\betavec = (\beta_1, \dots, \beta_Q)\tran$, the first order difference is $\beta_{q} - \beta_{q-1}, q = 2, \dots, Q$. The penalty term is
\begin{align*}
	P(\lambda, \betavec) = \lambda\sum_{q=2}^{Q} (\beta_{q} - \beta_{q-1})^2.
\end{align*}
Then the least squares estimation is to minimize the penalized loss function,
\begin{align*}
	L(\betavec, \lambda)=(\yvec - \Xmat\betavec)\tran(\yvec - \Xmat\betavec) + P(\lambda, \betavec).
\end{align*}
With appropriate selection of the regularization parameter $\lambda$, the overfitting issue can be mitigated without compromising the quality of the fit. This balance ensures that the model is both flexible and robust. Despite the usefulness of the original P-splines, it is not without challenges. Specifically, tuning the parameter $\lambda$ can be difficult, and multiple criteria can be applied here, including the Akaike information criterion (AIC) and the Bayesian information criterion (BIC). \citeN{Lang2004} adapted the original P-splines by incorporating a Bayesian perspective, leading to the Bayesian P-splines. In the Bayesian P-splines, the first order differences are assigned normal random walk priors as follows,
\begin{align*}
	\beta_{q} - \beta_{q-1} \sim \normal(0, \lambda^2), \quad q=2, \dots, Q.
\end{align*}
The penalty parameter $\lambda^2$ is assigned a non-informative inverse-Gamma (IG) distribution, denoted as $\IG(a,b)$, with small $a$ and $b$, which reflects minimal prior information about the parameter. Subsequently, the posterior likelihood can be derived. Sampling scheme, such as MCMC, can be used to explore the inferential properties of the model.

In the context of model~\eqref{eqn:GPM.model.spline.re}, for the vector of coefficients $\betavec_{jm}$, $j=1,\dots,J$, $m=1,\dots,M$, we denote its first order difference vector as $\uvec_{jm} = (u_{jm1}, \dots, u_{jm,Q_m-1})\tran$,
where
\begin{align*}
	u_{jm1} = \beta_{jm2} - \beta_{jm1}, \dots,u_{jm,Q_m-1} = \beta_{jmQ_m} -\beta_{jm,Q_m-1}.
\end{align*}
The random walk prior for any $u_{jmq}$, $q=1,\dots,Q_m-1$ with penalty parameter $\lambda_{m}$ is
\begin{align*}
	\pi(u_{jmq}) \propto \normal(0, \lambda_{m}^2),\quad \pi(\lambda_{m}^2) \propto \IG(a, b),
\end{align*}
where the hyper-parameter $a$ and $b$ can be assigned with small values such as, $a=1$, $b=0.005$.

\subsection{Monotone and Convex Shape Constraints} \label{subsec:shape.constraint}

Various studies have been conducted to ensure the shape of spline fitting (e.g., \citeNP{Ramsey1988}, and \citeNP{Meyer2008}). Despite these advancements, overfitting remains a challenge, and the selection of knots and degrees requires additional efforts. \citeN{Pya2015} proposed a method that employs reparametrization of the P-splines coefficients to generate a variety of shapes. From a Bayesian perspective, multi-layer transformations involving a large number of parameters can result in a highly complex posterior geometry space. This complexity makes it challenging for MCMC samplers to traverse the space thoroughly.

In this paper, we integrate the ideas of \citeN{Lang2004} and \citeN{Pya2015} to design a penalty scheme that penalizes regions where the shape does not conform to the desired characteristics. This approach seeks a balance between preserving the intended shape while allowing for adequate flexibility in the model.

To convey the proposed idea, we first need to discuss shape constraints. For a continuous function $\ffun(x)$, assuming its first and second derivatives $\ffun^{(1)}(x)$ and $\ffun^{(2)}(x)$ exist, if $\ffun^{(1)}(x) \geq 0$ for all $x$, then $\ffun(x)$ is non-decreasing. Conversely, if $\ffun^{(1)}(x) \leq 0$ for all $x$, then $\ffun(x)$ is non-increasing. Further if $\ffun^{(2)}(x) \leq 0$ for all $x$, then $\ffun(x)$ is concave, and if $\ffun^{(2)}(x) \geq 0$ for all $x$, then $\ffun(x)$ is convex. Such a function $\ffun(x)$ can be approximated using linear combinations of B-spline bases. We denote the non-negative B-spline basis functions of degree $r$ as $B_{q,r}(x)$, where $q=1,\dots,Q$. With the corresponding coefficient vector $\betavec$, $f(x) = \sum_{q=1}^{Q} B_{q,r}(x) \beta_{q}$.
Suppose we use equal-spaced knots here and the distance between knots is $z$. Based on the detailed calculations of B-splines derivatives in \citeN{Deboor1972}, the first and second order derivatives can be derived as,
\begin{align*}
	&\ffun^{(1)}(x) = \frac{1}{z} \sum_{q=2}^{Q}B_{q,r-1}(x) (\beta_{q} - \beta_{q-1}),\\
	&\ffun^{(2)}(x) = \frac{1}{z^2} \sum_{q=3}^{Q}B_{q,r-2}(x) (\beta_{q} - 2\beta_{q-1} + \beta_{q-2}).
\end{align*}
The necessary condition for $\ffun(x)$ to be non-decreasing is $\beta_q - \beta_{q-1} \geq 0$ for $q \geq 2$. For $\ffun(x)$ to be convex, it requires $\beta_{q} - 2\beta_{q-1} + \beta_{q-2} \geq 0$ for $q \geq 3$.
Similar conditions can be derived for non-increasing and concave cases. From a Bayesian perspective, these constraints can be incorporated by assigning informative priors to the corresponding parameters.

Recall the first order difference for $\betavec_{jm}$ is $\uvec_{jm}$, and $\ffun_{jm}(x) = \sum_{q=1}^{Q_m} B_{mq}(x) \beta_{jmq}$,
where $B_{mq}(x)$ is a B-spline basis function with degree 3.
To ensure the increasing shape of $\ffun_{jm}(x)$, we introduce a truncated normal prior $\trunnor(0, \lambda_{m}^2)$ (\shortciteNP{Evans2011}) with lower bound 0, variance $\lambda_{m}^2$ for all the elements in $\uvec_{jm}$ as follows,
\begin{align*}
	\pi(u_{jmq}) \propto \trunnor(0, \lambda_{m}^2),
\end{align*}
where $u_{jmq} \geq 0$. Consequently, this ensures that all samples drawn from the prior are non-negative, resulting in the desired increasing shape of $\ffun_{jm}(x)$.

Intuitively, we can apply a similar scheme to achieve a convex shape by assigning a truncated normal prior to the second order difference. This approach will result in the following term,
\begin{align*}
	\beta_{jmq} - 2\beta_{jm,q-1} + \beta_{jm,q-2} = u_{jmq} - u_{jm,q-1} \sim \trunnor(0, \lambda_m^2),
\end{align*}
for $q=3,\dots, Q_m$. However, this approach encounters difficulties due to the involvement of two additional layers of reparametrization. The substantial increase in the number of parameters generated by reparametrization leads to a couple of challenges. First, multi-layer reparametrization causes the posterior geometry space for $\betavec_{jm}$ to become highly complex, making it difficult for MCMC samplers to traverse. Second, in cases where the data provide little or no information about the shape of $\ffun_{jm}(x)$, the excessive number of parameters created by the additional layers can be redundant and highly correlated. This often results in pathological behavior of MCMC samplers in practice.

Alternatively, based on the Bayesian P-splines setting $u_{jmq}  \sim \normal(0, \lambda_{m}^2)$, we devise a shape-constrained penalty term ($\mbox{SCP}$) to address the parts of the function that mismatch the desired shape. This helps ensure that the model follows the shape constraints more closely without complexity or reparametrization issues. We introduce a positive shape constraint parameter $\delta_m$ and construct the SCP as proportional to an exponential quadratic form as follows:
\begin{align*}
	\mbox{SCP}(\betavec_{jm},\delta_m) & \propto \prod_{q=3}^{Q_m}\exp\left[-\frac{(\beta_{jmq} - 2\beta_{jm,q-1}+ \beta_{jm,q-2})^2}{2\delta_m^2} \onevec\left\{\beta_{jmq} - 2\beta_{jm,q-1} + \beta_{jm,q-2} < 0\right\}\right] \\
	& = \prod_{q=2}^{Q_m-1}\exp\left[-\frac{(u_{jmq} - u_{jm,q-1})^2}{2\delta_m^2} \onevec\left\{u_{jmq} - u_{jm,q-1} < 0\right\}\right].
\end{align*}
The above term penalizes the part of the function $\ffun_{jm}(x)$ that is concave, which is characterized by a positive second order difference. By applying the $\mbox{SCP}$, the model is encouraged to adhere more closely to the desired shape constraints.
In cases where the data lack sufficient information to determine the shape of $\ffun_{jm}(x)$,  we should use informative prior on $\delta_m$ rather than non-informative priors. In our application, the prior choice for the positive parameter $\delta_m$ is $\pi(\delta_m) \sim \normal(0, \eta^2)$.
Here $\eta$ is a fixed global constant which is determined by users, we suggest using small values such as 0.05 or 0.01 to provide strong shape constraint.

\subsection{Posterior Inference}\label{subsec:postinfer}
In this section, we introduce the Bayesian framework for the degradation model in Section~\ref{subsec:model}. With a sufficient number of samples, we can derive estimates, standard deviations (SDs), and credible intervals (CIs) for the parameters. In this paper, we employ the No-U-Turn Sampler (NUTS) (\citeNP{Hoffman2011}) that built in the state-of-art statistical modeling platform \text{STAN} (\citeNP{STAN2018}). Without loss of generality and for simplicity in discussion, we consider two covariates $M=2$ in the model. We assume that the true effect function for covariate 1 is non-decreasing and for covariate 2 is convex. Essentially we consider two types of effect functions. When $M>2$, the modeling framework can be easily extended.

Recall the general path model in Section~\ref{subsec:model} is
\begin{align*}
	y_{ijk} = \Dfun(\tau_{ijk};\thetavec_j, w_{ij}) + \epsilon_{ijk}.
\end{align*}
Let $\Dfun_{ijk}$ denote $\Dfun(\tau_{ijk};\thetavec_j, w_{ij})$ and $\sigma_l = \log \left(\sigma\right)$. For all $i$, $j$, $k$, the Bayesian hierarchy framework is
\begin{align*}
	& y_{ijk} \sim \normal\left(\Dfun_{ijk}, \sigma^2\right), \text{ } \pi(\alpha_j) \propto \pi(\gamma_j) \propto 1, \text{ } \pi\left(\sigma_l\right) \propto 1, \\
	& \pi(\Omegamat) \propto \mbox{LKJ}(\zeta),\text{ } \pi(\sigma_{j}) \propto \mbox{HCAU}(a).
\end{align*}
Here, $\mbox{HCAU}(a)$ denotes the half-Cauchy distribution with scale parameter $a$, and choices for $a$ can be determined using prior predictive checks. In this paper, we set $a=1$. We employ the $\mbox{LKJ}(\zeta)$ distribution with parameter $\zeta = 1$ as the prior for the correlation matrix $\Omegamat$, which is defined by \citeN{Lewandowski2009}. When $\zeta = 1$, each $\rho_{jj'}, j\neq j'$ in $\Omegamat$ is uniformly distributed among the interval $[-1,1]$.

Employing the Bayesian P-splines approach as we discussed earlier in this section, for the $j$th DC, $j=1,\dots,J$, the prior setting for $X_1$'s coefficients $\betavec_{j1}$ is,
\begin{align*}
	& \beta_{j11} \sim \pi(\beta_{j11}), \text{ } \pi(\uvec_{j1}) \propto \trunnor(0, \lambda_{1}^2\Imat), \text{ } \pi(\lambda_1^2) \propto \IG(a_1, b_1),
\end{align*}
where $\Imat$ is the identity matrix, and $a_1$ and $b_1$ are non-informative fixed parameters for IG distribution. Common choices for $a_1$ and $b_1$ are 1, 0.005.
The prior for $\beta_{j11}$ is denoted as $\pi(\beta_{j11})$. In practice, we recommend using a flat prior during the early stage of model diagnostics. Using non-informative priors, such as flat prior can help reveal any issues with the model structure or data without the influence of strong prior assumptions.

Once the model diagnostics have been performed and potential issues addressed, users can replace the flat prior with more informative priors that concentrate in the ideal range. These informative priors can incorporate domain-specific knowledge or insights from previous diagnose, which can improve the efficiency of the MCMC sampling process. For example, the flat priors for $\alpha_j$ and $\gamma_j$ can be replaced with uniform distributions with specified upper and lower bounds. The flat prior for $\beta_{j1}$ can be replaced with normal distribution. Here we can derive the joint prior for $\betavec_{j1}$ as,
\begin{align*}
	\pi(\betavec_{j1}) \propto \pi(\beta_{j11})\pi(\uvec_{j1})\pi(\lambda_{1}^2).
\end{align*}

The prior for $\betavec_{j2}$ is slightly different from that of $\betavec_{j1}$, as its shape  is convex. This requirement involves the use of the $\mbox{SCP}$ to ensure the desired shape. The prior setting for $\betavec_{j2}$ is
\begin{align*}
	& \beta_{j21} \sim \pi(\beta_{j21}), \text{ } \pi(\uvec_{j2}) \propto \normal(0, \lambda_{2}^2\Imat ) \cdot \mbox{SCP}(\betavec_{j2},\delta),\\
	& \pi(\lambda_{2}^2) \propto \IG(a_2,b_2), \text{ } \pi(\delta) \propto \normal(0, \eta^2),
\end{align*}
where $\pi(\beta_{j21})$ is the prior for $\beta_{j21}$. The term $\normal(0, \lambda_{2}^2\Imat ) \cdot \mbox{SCP}(\betavec_{j2},\delta)$ is the pdf of the normal distribution $\normal(0, \lambda_{2}^2\Imat)$ multiplied by the $\mbox{SCP}$.

Similar to $\betavec_{j1}$, the joint prior for $\betavec_{j2}$ is,
\begin{align*}
	\pi(\betavec_{j2}) \propto \pi(\beta_{j21})\pi(\uvec_{j2})\pi(\lambda_{2}^2)\pi(\delta).
\end{align*}
We denote $\thetavec=\left( \thetavec_1\tran, \dots, \thetavec_J\tran\right)\tran$,
and $\wvec = (\wvec_{1}\tran, \dots, \wvec_{I}\tran)\tran$.
Let $\yvec_{j} = \left(y_{1j1}, \dots, y_{1jn_1}, \dots, y_{Ijn_I} \right)\tran$, and $\yvec = \left(\yvec_{1}\tran, \dots, \yvec_{J}\tran\right)\tran$. The joint likelihood conditional on fixed parameters and random effects is,
\begin{align*}
	\likh\left(\yvec | \thetavec, \wvec, \sigma_l\right) = \prod_{i,j,k} \Phi\left[\frac{y_{ijk}-\Dfun_{ijk}}{\exp\left(\sigma_l\right)}\right] \prod_{i} |\Sigma|^{-\frac{1}{2}} \exp\left(-\frac{1}{2} \wvec_{i}\tran \Sigma^{-1} \wvec_i\right),
\end{align*}
where $\Phi(\cdot)$ stands for the standard normal pdf.

The posterior likelihood $\likh(\thetavec, \wvec, \Sigmat, \sigma_l| \yvec, \Xmat)$ is proportional to the joint likelihood multiply by priors,
\begin{align*}
	\likh\left( \thetavec, \wvec, \Sigmat, \sigma_l| \yvec\right) \propto \likh(\yvec|\thetavec, \wvec) \pi(\thetavec)\pi(\Sigmat),
\end{align*}
where
\begin{align*}
	\pi(\thetavec) \propto \prod_{j}\left[\pi(\alpha_j)\pi(\gamma_j)\pi(\betavec_{j1})\pi(\betavec_{j2})\right], \text{ } \pi(\Sigmat) \propto \prod_{j}\pi(\sigma_j) \cdot \pi(\Omegamat).
\end{align*}
Based on the posterior likelihood, the log-posterior likelihood can be derived accordingly, which is,
\begin{align*}
	\loglikh(\thetavec, \wvec) &= \sum_{j}\left\{\log\left[\pi(\alpha_j)\right] + \log\left[\pi(\gamma_j)\right] + \log\left[\pi(\betavec_{j1})\right] + \log\left[\pi(\betavec_{j2})\right] + \log\left[\pi(\sigma_j)\right]\right\} \\
	& + \log\left[ \likh(\yvec|\thetavec, \wvec)\right] + \log\left[ \pi(\Omegamat)\right] + \mbox{const}.
\end{align*}

\subsection{Failure Time Distribution Estimation}\label{subsec:ftdest}
The subsequent reliability analysis requires a model for the covariate processes. We adapt the model for covariate processes from \shortciteN{Hong2015}. To make the paper self-contain, we provide the details of the covariate model in Section~\ref{sec:modelcovprocess}.

For the $j$th DC, $j=1,\dots, J$, we say a DC-wise soft failure happens when the degradation path $\Dfun_j(t)$ first arrives at the DC-wise failure threshold $\Dfun_{f_j}$. As previously discussed, $\Dfun_j(t)$ is modeled as a function form of covariate process $\Xmat(t)$, with DC-wise parameters $\thetavec_j$, $j=1,\dots, J$, and random effect $w$. For convenience of illustration, we denote the degradation path as $\Dfun_j\left(t, \thetavec_j, w\right)$ in this section. We denote the DC-wise failure time random variable as $T_j$, $j=1,\dots,J$. We say a specimen is failed when any DC-wise soft failure happens for the first time. The distribution of failure time of the system given parameters $\thetavec = (\thetavec_1\tran, \dots, \thetavec_J\tran)\tran$, at time $t$ is
\begin{align*}
	F(t, \thetavec)  &= 1 - \Pr \left(T_1 \geq t, \dots, T_J \geq t \right)\\
	& =\E_{\Xmat(t)} \E_{w}\left\{  1 - \Pr \left[\Dfun_1 (t;\thetavec_1, w) \leq \Dfun_{f_1}, \dots, \Dfun_J (t;\thetavec_J, w) \leq \Dfun_{f_J} \right]\right\}.
\end{align*}
Utilizing an estimator for the degradation path allows for the estimation of $F(t,\thetavec)$.

To estimate the degradation path, we not only need estimators for parameters $\thetavec$, and random effects $w$, we also need estimator for covariate process $\hat{\Xmat}(t)$, which can be obtained from the model for the covariate process. Consequently, the posterior samples of $F(t, \thetavec)$ can be drawn by the following Monte Carlo based algorithm, with $(\thetavec^{\ast}, \Sigmat^{\ast})$ denoting a posterior sample of $(\thetavec, \Sigmat)$,
\begin{enumerate}
	\item[\textbf{Algorithm 1}]
	\item[\textbf{Step 1:}] Simulate $\hat{\Xmat}(t)$.
	\item[\textbf{Step 2:}] Generate $\wvec \sim \normal(\zerovec, \Sigmat^{\ast})$.
	\item[\textbf{Step 3:}] Compute failure time $t_f = \mbox{min}\left\{t: \Dfun\left(t, \thetavec_j^{\ast}, \wvec\right) \geq \Dfun_{f_j} \text{ for} j=1,\dots,J \right\}$.
	\item[\textbf{Step 4:}] Repeat above steps for $B$ times, we denote the failure time for loop $b$ as $t_f^{b}$, $b=1, \dots, B$.
	\item[\textbf{Step 5:}] Compute a sample from $F(t, \thetavec)$, $$F^{\ast}(t, \thetavec) = \frac{1}{B} \sum_{b=1}^{B}\indi \left\{t_{f}^{b} \leq t\right\}.$$
\end{enumerate}

Traversing all posterior samples of $(\thetavec,\Sigmat)$, we achieve a posterior distribution for $F(t,\thetavec)$. The estimator for $F(t,\thetavec)$, say $\hat{F}(t, \thetavec)$ is the median of the distribution. To obtain a $95\%$ CI for $\hat{F}(t,\thetavec)$, we repeat Algorithm 1 for all the bootstrap samples of $\thetavec_X$ and select the $2.5\%$ and $97.5\%$ quantiles of the results as CI bounds.

\section{Simulation Study}\label{sec:simulations}
A simulation study is designed to assess the proposed method. This study simulates datasets that replicate the structure of the weathering data. Our primary focus is to investigate the performance of the estimators of parameters, the effect functions, and the failure time distributions.

\subsection{Simulation Settings}\label{subsec:simset}
For simplicity, we limit the number of dynamic covariates to two: $\Xmat_1$ and $\Xmat_2$. Their corresponding effect functions are denoted as $\ffun_{1}(x)$ and $\ffun_{2}(x)$, where $\ffun_{1}(x)$ is increasing and $\ffun_{2}(x)$ is convex. The details of covariate processes and effect functions are included in Section~\ref{sec:cov.proc.eff.sim}.

All specimens in this simulation are exposed to the covariates for 300 days and $J=2$ DCs are considered. We construct 27 scenarios on combinations of three variables: the number of specimens ($I=10, 20, 50$), the number of degradation measurements per specimen within the exposed period ($n_i=10, 20, 50$), and correlation among random effects of different DCs ($\rho=0.2, 0.5, 0.9$). The fixed parameters are set as $\alphavec = (\alpha_{1}, \alpha_{2})\tran = (0.3,0.5)\tran$, and $\gammavec = (\gamma_{1}, \gamma_{2})\tran = (1,1.1)\tran$.

For each scenario, we generate 500 datasets with SD of the error term $\sigma = 0.01$, and SDs for random effect $\sigma_j = 0.1$, $j=1,2$. The random effect $\wvec_i = (w_{i1}, w_{i2})\tran$, $i=1,\dots,I$, is set to have multivariate normal distribution $\wvec_i \sim \normal(\zerovec, \Sigmat)$, where
\begin{align*}
	\Sigmat = \Dmat\Omegamat\Dmat\tran = \begin{pmatrix}
		\sigma_1^2 & \rho\\
		\rho & \sigma_2^2
	\end{pmatrix}, \text{ }
	\Omegamat =
	\begin{pmatrix}
		1 & \rho\\
		\rho & 1
	\end{pmatrix}, \text{ } \Dmat = \mbox{Diag}(\sigma_{1}, \sigma_{2}).
\end{align*}
The failure threshold for this simulation is set as $-0.2$ for the first DC and $-0.45$ for the second DC. To highlight the significance of interdependent random effects among DCs, we compare reliability estimation results from two models based on different setting of correlation. The first one is called the correlated model, which assume $\rho \neq 0$. The second one is called the independent model, which assume $\rho = 0$.

For each simulated dataset, posterior samples are derived from four independent MCMC chains. Each chain consists of 1,000 iterations, including 200 iterations for burn-in. The Bayesian estimators for the parameters are the medians of these samples. We assess the distributions of estimators for $\gammavec$, and elements of $\Sigmat$ and $\Omegamat$. Due to non-identifiability of $\alpha_j$, we obtain estimates for specimen level asymptotes $\alpha_j\exp(\omega_{ij})$, where $i=1,\dots, I$, and $j=1,\dots,J$.  For the evaluation of effect functions and reliability functions, the averaged Root Mean Square Error (Avg RMSE) provides a more insightful analysis. The Avg RMSE for a function $f(x)$ with given points of $x$ is calculated as
$$\text{Avg RMSE} = \frac{1}{x_{\text{max}}-x_{\text{min}}}\int_{x_{\text{min}}}^{x_{\text{max}}}\mbox{RMSE}(u)du.$$

\subsection{Simulation Findings}\label{subsec:sim.res}
\begin{figure}
	\centering
	\includegraphics[width=0.65\linewidth]{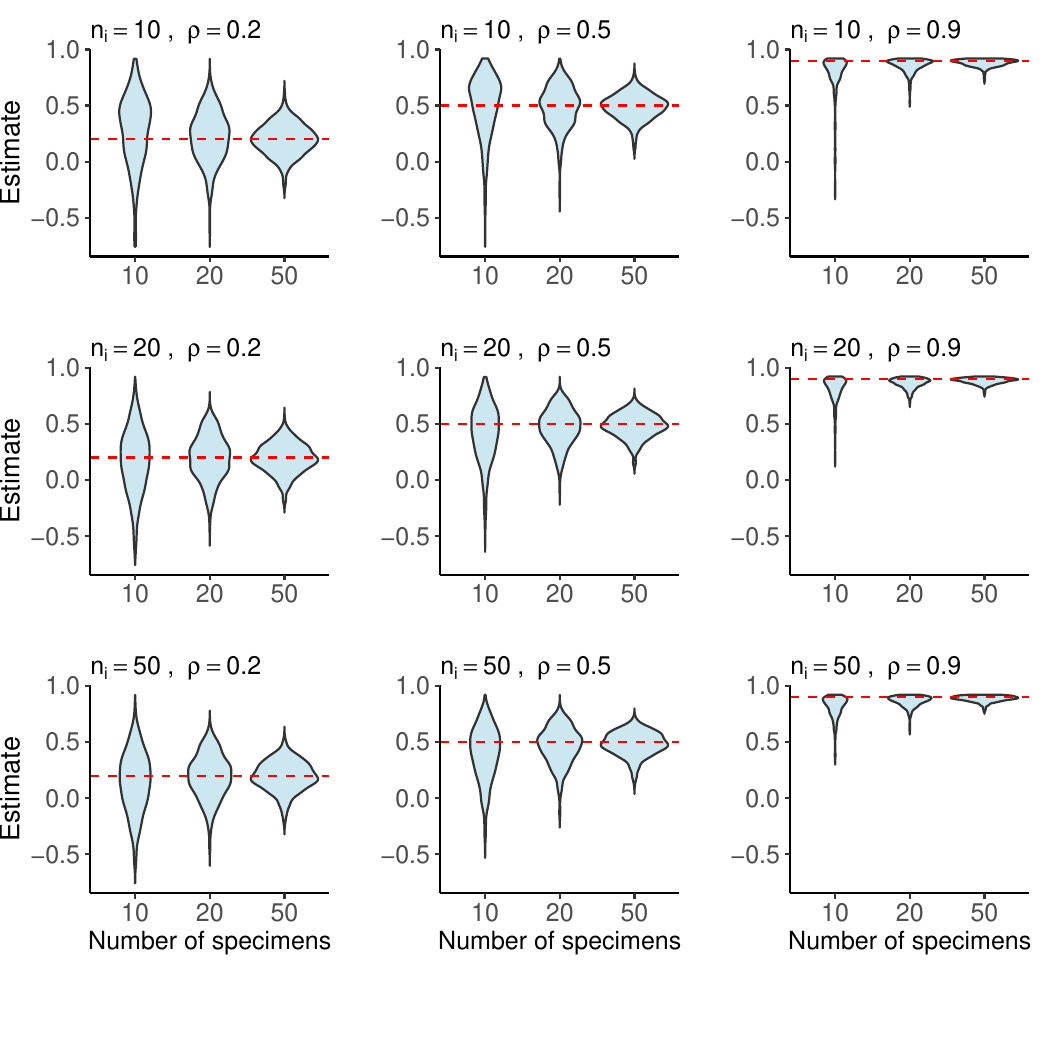}
	\caption{Violin plot for posterior distributions of $\rho$ under different combinations of $I$ and $n_i$. The red dashed line shows the true $\rho$.}
	\label{fig:sim.rho.plt}
\end{figure}

\begin{figure}
	\centering
	\includegraphics[width=0.65\linewidth]{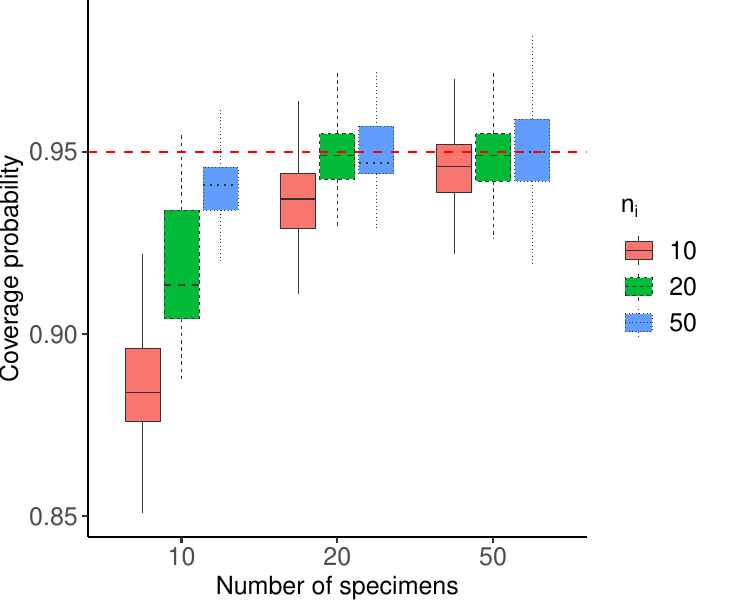}
	\caption{Boxplot of CPs of specimen asymptotes given combinations of number of specimens $I$ and $n_i$ for DC 1. The dashed horizontal line shows the nominal $95\%$ CP.
	}
	\label{fig:sim.asymptote_cp.plt}
\end{figure}

Table~\ref{tab:sim.par.inf} presents the distribution statistics of posterior medians for model's parameters of the first DC, including $\gamma_1$, $\sigma_1$, and the SD for the error term $\sigma$. The ``Median'' column in Table~\ref{tab:sim.par.inf} represents the mean of posterior medians obtained from simulated datasets. The results indicate that both an increase in the number of specimens ($I$) and in the number of measurements per specimen ($n_i$) contribute to improved estimations by reducing bias and SD.  Overall, the coverage probability (CP) is close to the nominal 95\%, except for when $n_i$ is small. The conclusions for $\gamma_2$ and $\sigma_{2}$ are similar and presented in Table~\ref{tab:supp.sim.par.inf2}. Figures~\ref{fig:supp.sim.r1.plt},~\ref{fig:supp.sim.r2.plt},~\ref{fig:supp.sim.sig_w1.plt}, and~\ref{fig:supp.sim.sig_w2.plt} show  violin plots that display the distribution of estimates for $\gamma_{1}$, $\gamma_{2}$, $\sigma_{1}$, and $\sigma_{2}$, respectively. The results are consistent with the conclusions based on the tables.
Figure~\ref{fig:sim.rho.plt} shows the estimation results for correlation parameter $\rho$. When $n_i$ is fixed, increasing $I$ can shorten the tails and enlarge the areas near true value, indicating improved estimation. However, increasing the $n_i$ when $I$ is fixed can not lead to the same conclusion.
In addition, Figure~\ref{fig:sim.asymptote_cp.plt} shows that increasing $I$ or $n_i$ can improve the CPs for catching specimen level asymptotes.

\begin{table}
	\centering
	\caption{Summary of posterior medians for a subset of parameters under different scenarios.}\label{tab:sim.par.inf}
	\begin{tabular}{ccccccc}\toprule
		$I$ & $n_i$ & Parameter & Median & Bias$\times10^{2}$ & SD$\times10^{2}$ & CP\\
		\midrule
		10 &  10 &  $\gamma_1$ & 0.970 & 6.480 & 9.792 & 0.906 \\
		&  20 &            & 0.980 & 4.256 & 4.994 & 0.960 \\
		&  50 &            & 0.991 & 2.751 & 3.291 & 0.946 \\
		\midrule
		20 &  10 & $\gamma_1$ & 0.977 & 4.051 & 4.388 & 0.944 \\
		&  20 &            & 0.988 & 2.667 & 3.159 & 0.953 \\
		&  50 &            & 0.994 & 1.804 & 2.191 & 0.946 \\
		\midrule
		50 &  10 &  $\gamma_1$ & 0.990 & 2.402 & 2.800 & 0.957 \\
		&  20 &            & 0.995 & 1.655 & 2.024 & 0.947 \\
		&  50 &            & 0.998 & 1.120 & 1.357 & 0.949 \\
		\midrule
		10 &  10 & $\sigma$ & 0.010 & 0.056 & 0.062 & 0.898 \\
		&  20 &          & 0.010 & 0.034 & 0.039 & 0.942  \\
		&  50 &          & 0.010 & 0.019 & 0.024 & 0.943 \\
		20 &  10 & $\sigma$ & 0.010 & 0.031 & 0.037 & 0.939 \\
		&  20 &          & 0.010 & 0.021 & 0.025 & 0.949 \\
		&  50 &          & 0.010 & 0.013 & 0.017 & 0.944 \\
		50 &  10 & $\sigma$ & 0.010 & 0.019 & 0.024 & 0.952 \\
		&  20 &          & 0.010 & 0.013 & 0.016 & 0.959 \\
		&  50 &          & 0.010 & 0.008 & 0.010 & 0.956 \\
		\midrule
		10 &  10 &  $\sigma_{1}$ & 0.100 & 1.990 & 2.501 & 0.897\\
		&  20 &          & 0.099 & 1.850 & 2.351 & 0.936 \\
		&  50 &          & 0.100 & 1.969 & 2.457 & 0.939 \\
		20 &  10 & $\sigma_{1}$ & 0.099 & 1.307 & 1.609 & 0.938 \\
		&  20 &         & 0.098 & 1.206 & 1.492 & 0.944 \\
		&  50 &         & 0.101 & 1.297 & 1.623 & 0.949 \\
		50 &  10 & $\sigma_{1}$ & 0.099 & 0.812 & 1.020 & 0.959 \\
		&  20 &          & 0.100 & 0.789 & 0.980 & 0.944 \\
		&  50 &          & 0.100 & 0.800 & 0.991 & 0.949 \\
		\bottomrule
	\end{tabular}
\end{table}

\begin{figure}
	\centering
	\includegraphics[width=0.8\linewidth]{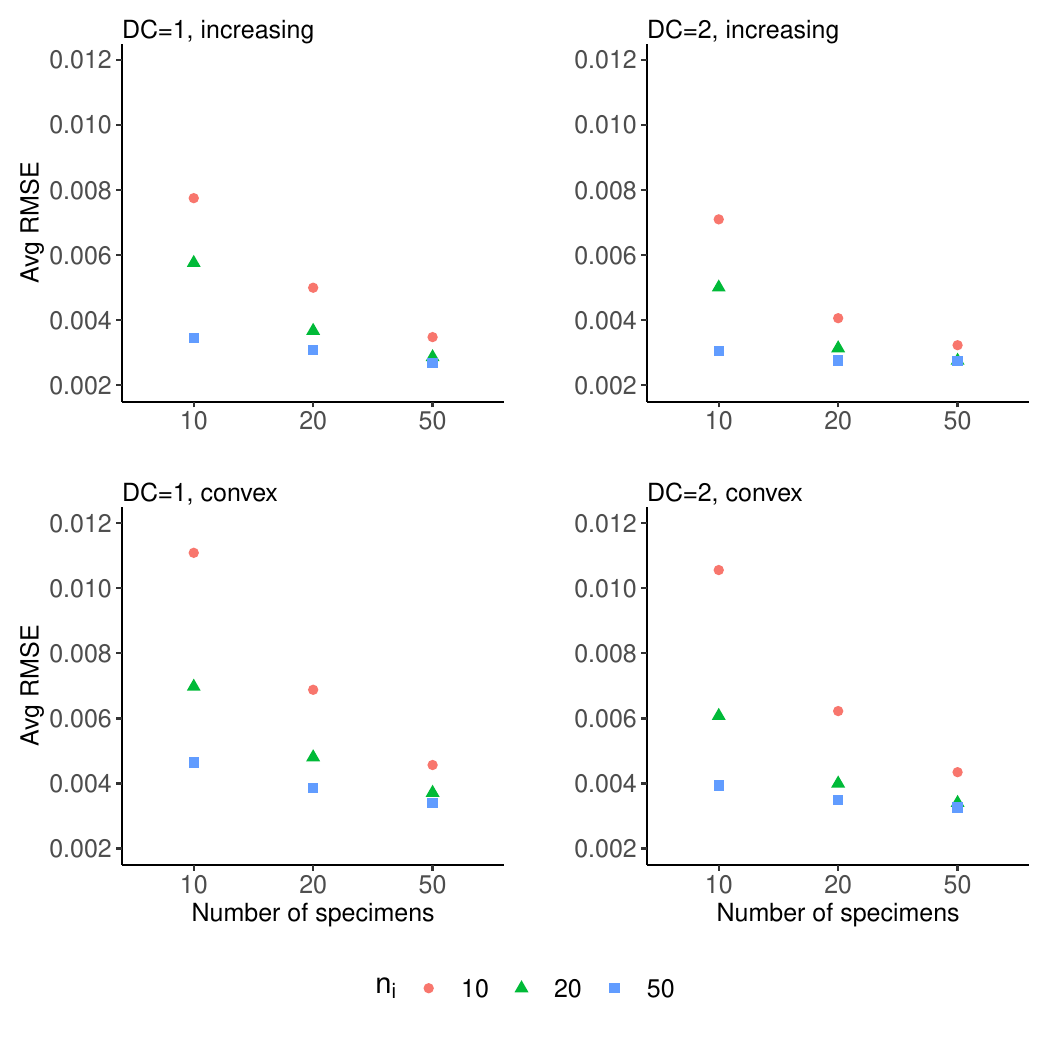}
	\caption{Avg RMSE plot of estimators of effect functions given different shapes, DCs, and scenarios.
	}
	\label{fig:sim.effect_avg_rmse.plt}
\end{figure}

Regarding the estimators for effect functions, Figure~\ref{fig:sim.effect_avg_rmse.plt} shows the Avg RMSE values, which decline with an increase in $I$ or $n_i$. Within the same DC, the Avg RMSE for the increasing shape is consistently lower than that for the convex shape, suggesting a more precise estimation for the increasing shape. With the increase of $I$, the differences in Avg RMSE among different $n_i$ values diminish, suggesting that we are approaching optimal estimation performance. Figures~\ref{fig:supp.sim.effect_cp1.plt} and \ref{fig:supp.sim.effect_cp2.plt} illustrate the CP of the effect function given the range of covariate values, demonstrating accurate effect function estimations with CPs close to 95\%. An increase in $I$ and $n_i$ improves the CPs for both shapes, with the increasing shape consistently outperforming the convex shape, being consistent with the Avg RMSE findings. Figures~\ref{fig:supp.sim.effect_inc_1_coverage.plt},~\ref{fig:supp.sim.effect_inc_2_coverage.plt},~\ref{fig:supp.sim.effect_convex_1_coverage.plt}, and~\ref{fig:supp.sim.effect_convex_2_coverage.plt} show the ribbon plots of effect functions.  As increase of $I$ or $n_i$, the ribbons become narrow and concentrate to the solid lines.

\begin{figure}
	\centering
	\includegraphics[width=0.8\linewidth]{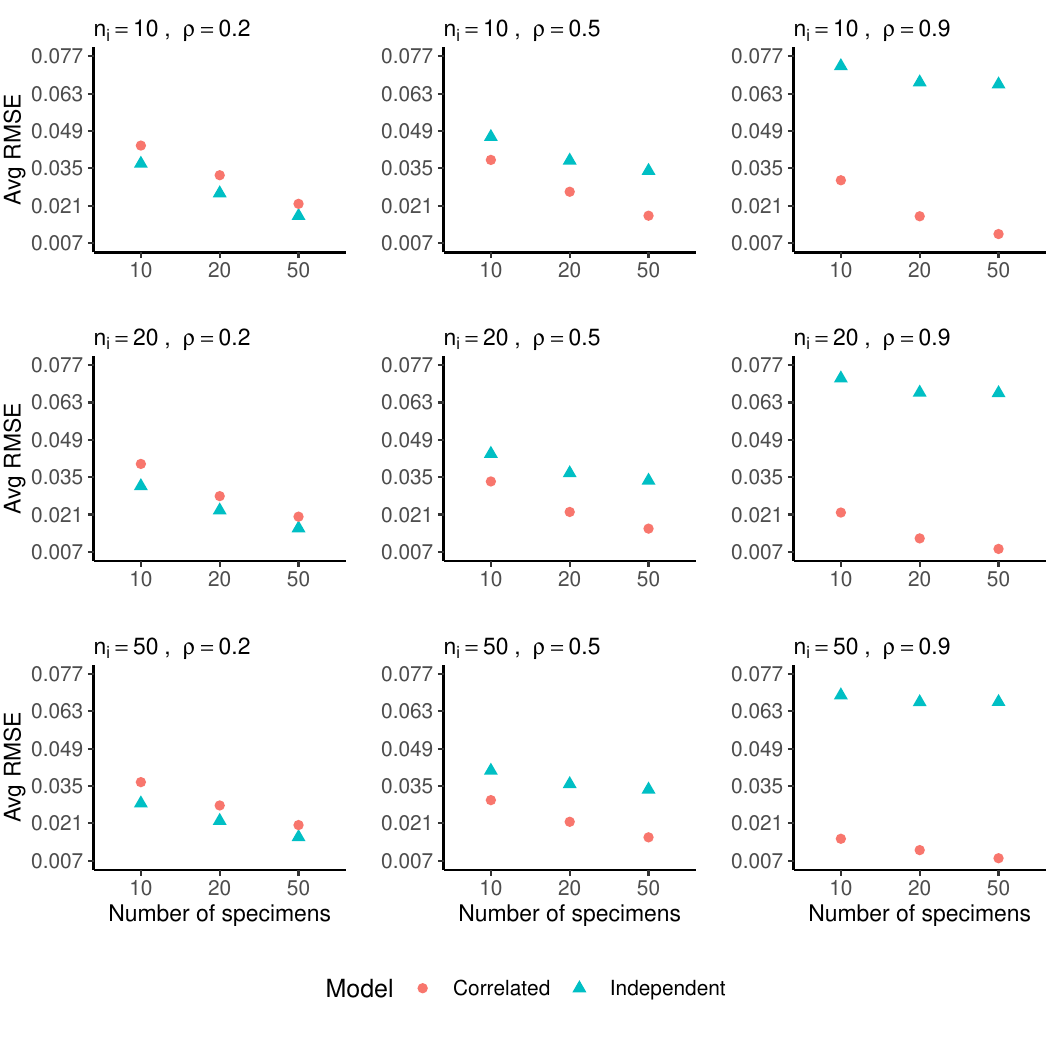}
	\caption{Avg RMSE for estimators of reliability function under different scenarios.
	}
	\label{fig:sim.reliability.rmse.box.plt}
\end{figure}

Figure~\ref{fig:sim.reliability.rmse.box.plt} shows a comparison of Avg RMSE between the correlated model and the independent model regarding to reliability estimation. The independent model can achieve better estimation when correlation is weak due to a smaller Avg RMSE. Conversely, the correlated model obtains significantly better estimation compared to the independent model for median to high correlations. Additionally, increasing $I$ improves reliability estimation in both models, given fixed $n_i$ and $\rho$. Figures~\ref{fig:supp.sim.dft_cp1.plt},~\ref{fig:supp.sim.dft_cp2.plt}, and~\ref{fig:supp.sim.dft_cp3.plt} display the CP of the failure time distribution from day 1 to day 150, confirming the conclusions discussed above. In conclusion, adopting the correlated model offers several advantages: it provides a clearer and more accurate representation of the system, results in a better estimation of failure time distributions, and ensures efficient use of resources by avoiding unnecessary early actions.

\section{Application to the Weathering Data}\label{sec:application}
In this section, we apply the model to the weathering data and obtain estimation for parameters, the effect functions, and model's reliability. According to the data, there are three DCs ($J=3$) and 36 specimens, denoted as $I=36$. Following \shortciteN{Lu2021}, the soft failure thresholds for each DC are $-0.40$, $-0.58$, and $-0.27$, respectively. Based on domain knowledge, we assume the effect functions are non-decreasing for UV and TEMP, and convex for RH. After pre-stage model tuning, the final priors for fixed effects $\alpha_j$, $\gamma_j$, $j=1,\dots,3$, are
\begin{align*}
	\pi(\alpha_j) \propto 1, \text{ } \pi(\gamma_j) \propto 1.
\end{align*}

\begin{table}
	\begin{center}
		\caption{Posterior distributions' medians for parameters, the corresponding posterior SDs, and the $95\%$ CIs.}\label{tab:deg.post.inf}
		\begin{tabular}{crrrrccrrrr}\toprule
			\multicolumn{5}{c}{Correlated Model}&&\multicolumn{5}{c}{Independent Model}\\\cline{1-5}\cline{7-11}
			\multirow{2}{*}{Para.}&\multirow{2}{*}{Median.} &\multirow{2}{*}{SD}&\multicolumn{2}{c}{ 95\% CI}&&
			\multirow{2}{*}{Para.}&\multirow{2}{*}{Median.} &\multirow{2}{*}{SD}&\multicolumn{2}{c}{ 95\% CI}\\\cline{4-5}\cline{10-11}
			&&& Lower & Upper&&&&& Lower & Upper\\\hline
			$\gamma_{1}$& 1.169 & 0.031 & 1.108    & 1.226 && $\gamma_{1}$& 1.163 & 0.031 & 1.103 & 1.224 \\
			$\gamma_{2}$& 1.068 & 0.008 & 1.053    & 1.082 && $\gamma_{2}$& 1.070 & 0.007 & 1.055 & 1.083 \\
			$\gamma_{3}$& 0.898 & 0.035 & 0.833    & 0.972 && $\gamma_{3}$& 0.889 & 0.035 & 0.824 & 0.958\\
			$\sigma_{1}$& 0.097 & 0.015 & 0.073    & 0.129 && $\sigma_{1}$& 0.093 & 0.014 & 0.068 & 0.123 \\
			$\sigma_{2}$& 0.117 & 0.015 & 0.092    & 0.152 && $\sigma_{2}$& 0.117 & 0.016 & 0.091 & 0.151 \\
			$\sigma_{3}$& 0.094 & 0.012 & 0.073    & 0.121 && $\sigma_{3}$& 0.094 & 0.012 & 0.072 & 0.122\\
			$\sigma$    & 0.016 & 0.000 & 0.016    & 0.016 && $\sigma$    & 0.016 & 0.000 & 0.015 & 0.016\\
			$\rho_{12}$ & 0.738 & 0.083 & 0.550    & 0.868 &&             &       &       &       & \\
			$\rho_{13}$ & 0.121 & 0.186 & $-$0.240 & 0.470 &&   		  &   	  &  	  & 	  & \\
			$\rho_{23}$ & 0.367 & 0.166 & 0.015    & 0.654 &&  			  &   	  &	      &		  & \\
			\bottomrule
		\end{tabular}
	\end{center}
\end{table}

We compare the estimations in failure time distribution of the correlated model and the independent model. In the independent model, the independence are obtained by assigning the correlation coefficients $\rho_{jj'}=0$, $j,j'=1,\dots, J$, $j\neq j'$. The number of splines bases we used is 60. To ensure convergence to our target distribution, we run 4 parallel chains, each with 5,000 iterations and the first 500-iteration for warm-up. Table \ref{tab:deg.post.inf} summarizes posterior distributions for parameters, including their estimates, which is the median.

\begin{figure}
	\centering
	\includegraphics[width=0.7\linewidth]{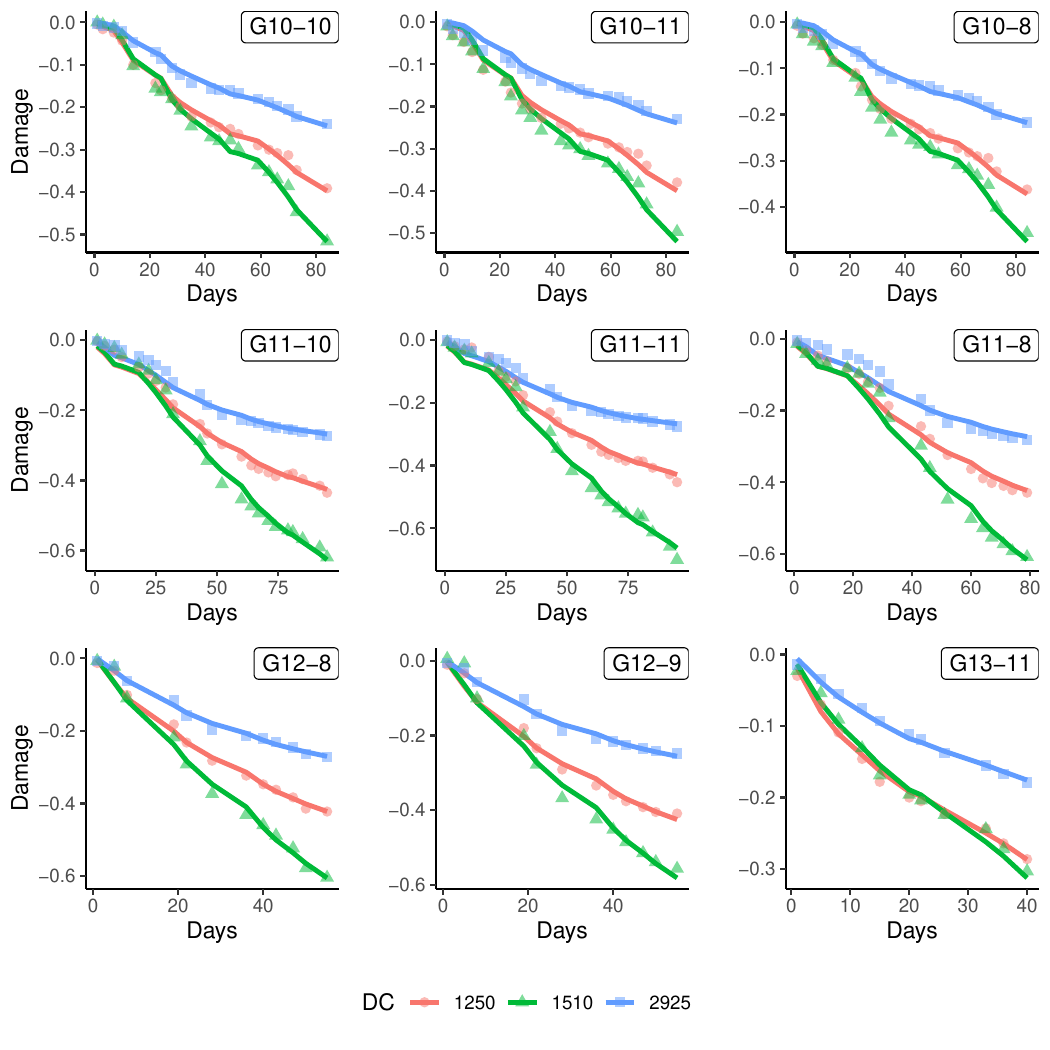}
	\caption{Fitted degradation paths for nine representative specimens. Each subfigure represents a specimen. }
	\label{fig:deg.path.fit.plt}
\end{figure}

The posterior distributions obtained using the correlated model and independent model are very similar except the correlated model has non-zero correlation coefficients $\rho_{12}$, $\rho_{13}$, and $\rho_{23}$. Therefore, we can expect that the fitted degradation path and the estimated covariate effect from the correlated model and the independent model are similar. Hereby we only show the results of the correlated model.

\begin{figure}
	\centering
	\includegraphics[width=0.7\linewidth]{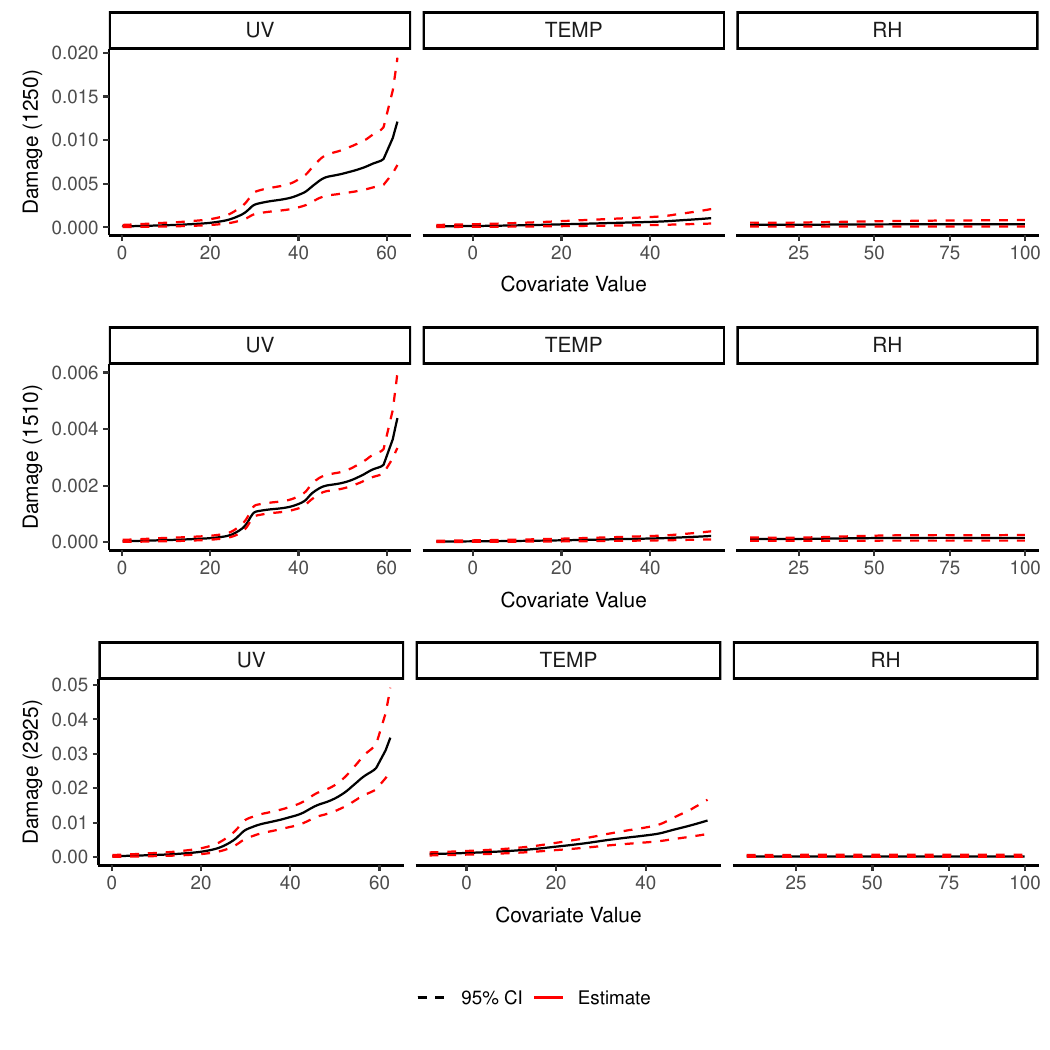}
	\caption{The estimated effect functions for UV, TEMP, and RH.}
	\label{fig:effect.plt}
\end{figure}
Figure~\ref{fig:deg.path.fit.plt} displays the fitted degradation path of selected specimens and we can see that our model fits the degradation path correctly. Figure~\ref{fig:effect.plt} shows the estimated covariate effects for different DCs with $95\%$ CIs. For same level exposure of covariates, the third DC (2925) receives more damage than the others. The shapes of the covariate effects are consistent with our assumptions. In all three types of degradation, UV has the largest effect compare with other two covariates. While RH has insignificant effect, which suggests removing it from future modeling. From Figure~\ref{fig:effect.plt}, the convexity of RH's effect is not as evident as the results from \shortciteN{Hong2015}. This is due to that the covariate information of RH is lack of sufficient variation. Figure~\ref{fig:cov.process}(c) shows that the range of RH is narrow and most of the data are concentrating on the range of $25$-$50$. In order to ensure the shape of effect to be what we assumed, more rigid splines can be employed, typically the ones in \shortciteN{Hong2015}. However, using rigid splines against the data will result in serious divergence problem in MCMC sampling as well as misleading inference conclusions. One major advantage of using Bayesian P-splines with shape constraints is that our splines are flexible enough under shape restrictions to express the information contained in the data.

\begin{figure}
	\centering
	\includegraphics[width=0.5\linewidth]{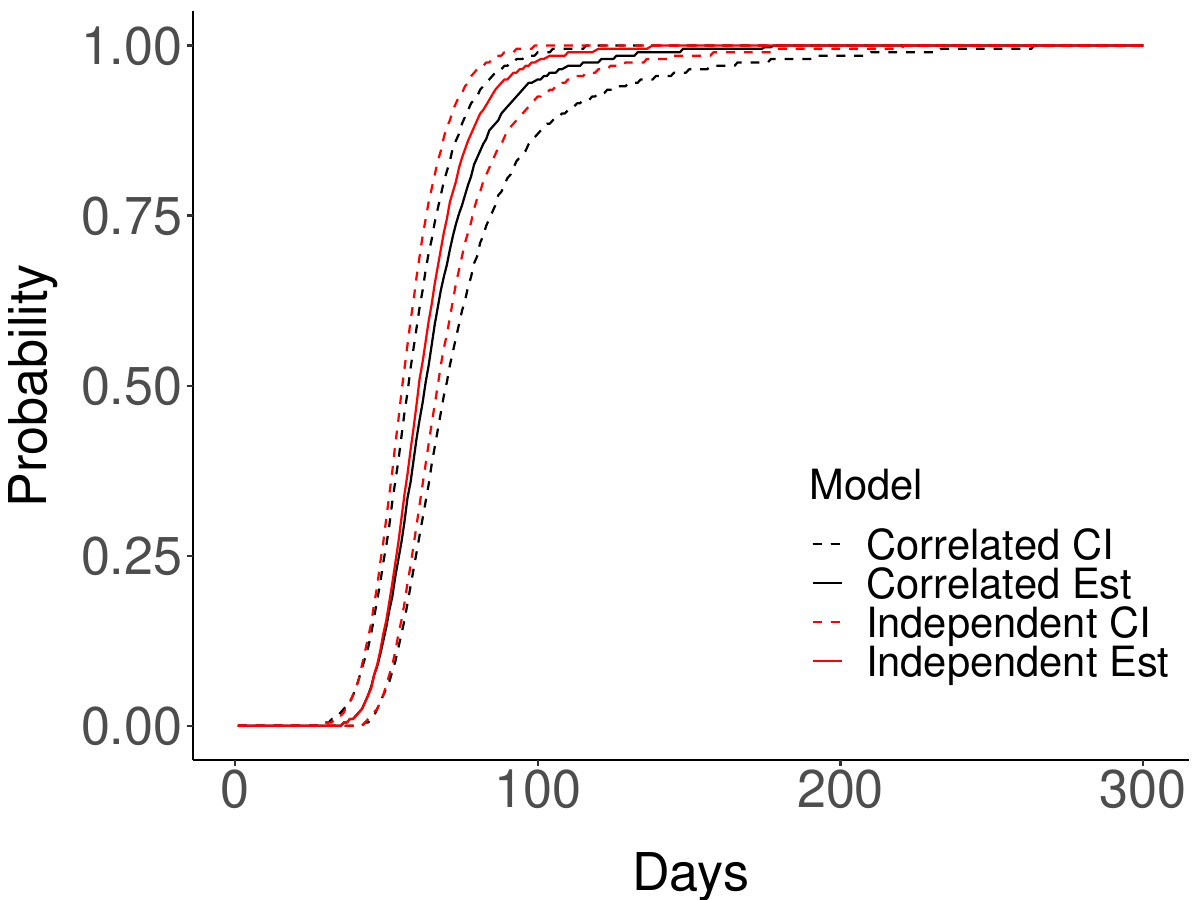}
	\caption{Comparison of predicted time distributions and CIs using Algorithm 1.}
	\label{fig:dft.plt}
\end{figure}
Figure \ref{fig:dft.plt} illustrates estimation of failure time distribution. We conducted a comparison between the distributions produced by the correlated and the independent models. The curve for the correlated model consistently lies below that of the independent model, indicating better reliability from the correlated model. The difference between the two estimated curves is not obvious. This supports our results from Table~\ref{tab:deg.post.inf}, showing that the estimated values of $\rho_{13}$ and $\rho_{23}$ are small. It also confirms the findings from the simulation study, which show a minimal difference in estimating the failure time distribution when correlations among DCs are weak.

\section{Conclusions and Areas For Future Research}\label{sec:conclusions}
To model multiple DC degradation processes influenced by dynamic covariates, we introduce a nonlinear mixed-effects model that accounts for environmental dynamic covariates. This model recognizes specimen variation using random effects and addresses correlations among multiple DCs through the covariance matrix of the random effects. The impact of dynamic covariates is described using shape-constrained Bayesian P-splines, adhering to domain-specific shape assumptions. Additionally, a Monte Carlo algorithm is designed to estimates the system's failure time distribution.

A simulation study is designed to evaluate model performance regarding to estimation of model parameters, effect functions, and reliability. The model parameters and the effect functions are estimated, and their accuracy improves with increased data dimensions. Our reliability estimation suggests the importance of incorporating correlation in modeling, with the correlated model providing more accurate system lifetime estimates than the independent model especially when correlation is median to strong. The method's efficacy is further demonstrated through the weathering data. In this application, an unstructured variance-covariance matrix is used to capture the correlation among DCs. The estimation results of correlation coefficients suggest that there exists stronger correlations between DCs (wavelength) of similar proximity.

In the future, our proposed method can be extended. By employing data-specific priors, we can develop a structured variance-covariance matrix. In addition, Bayesian autoregressive time series model can be applied for estimating covariate processes and can be included in the posterior likelihood of the model. Moreover, our approach is adaptable to other degradation applications requiring similar model configurations.

\section*{Acknowledgments}

The authors acknowledge the Advanced Research Computing program at Virginia Tech for providing computational resources. The work by Lin and Hong was partially supported by National Science Foundation Grant CMMI-1904165 to Virginia Tech. The work by Hong was partially supported  by the COS Dean's Discovery Fund (Award: 452021) and by the Data Science Faculty Fellowship (Award: 452118) at Virginia Tech.


\appendix

\section{Modeling Covariate Processes}\label{sec:modelcovprocess}
Section 6.2 of \shortciteN{Hong2015} introduces a general framework for modeling the multivariate covariate process with vector autoregressive time series model (VAR) based on \shortciteN{Reinsel1993} and estimation procedure for the model parameters. In this paper, we adopt the parametric model for estimating the covariate process developed by \shortciteN{Hong2015} because the covariates information used in \shortciteN{Hong2015} is same as the ones used in this paper. We provide a brief recap of the method in this section.

For the $m$th covariate, the model assumes that the covariate process is a multivariate Gaussian process with a mean function and a covariance function $a_m(t)$. The mean function is modeled as a summation of two terms, the trend  term $\mbox{Trend}_m(t)$, which describes the long-term trend, the seasonal term $\mbox{Seasonal}_m(t)$ which captures the periodic effect due to season change. The covariate process can be modeled as $x_m(t) = \mbox{Trend}_m(t) + \mbox{Seasonal}_m(t) + a_m(t)$.

Let $x_{1}(t)$, $x_{2}(t)$, and $x_3(t)$ denote the covariate values for UV, RH, and TEMP at time $t$, the covariate process model is
\begin{align*}
	\renewcommand{\arraystretch}{1}
	\begin{bmatrix}
		x_1(t) \\
		x_2(t) \\
		x_3(t)
	\end{bmatrix} = \begin{bmatrix}
		\mu_1 + \kappa_1 \sin \left[ \frac{2\pi}{365}(t-\eta_1)\right]\\
		\mu_2 + \kappa_2 \sin \left[ \frac{2\pi}{365}(t-\eta_2)\right]\\
		\mu_3 + \kappa_3 \sin \left[ \frac{2\pi}{365}(t-\eta_3)\right]\\
	\end{bmatrix} + \begin{bmatrix}
		\left(1 + \upsilon_1 \left\{1+\sin\left[\frac{2\pi}{365}(t-\zeta_1)\right]\right\}\right)\epsilon_1(t)\\
		\left(1 + \upsilon_2 \left\{1+\sin\left[\frac{2\pi}{365}(t-\zeta_2)\right]\right\}\right)\epsilon_2(t)\\
		\epsilon_3(t)
	\end{bmatrix}.
\end{align*}
The autocorrelation and correlation is modeled as VAR(2) as follows,
\begin{align*}
	\begin{bmatrix}
		\epsilon_1(t) \\
		\epsilon_2(t) \\
		\epsilon_3(t)
	\end{bmatrix} = \Phimat_1 \begin{bmatrix}
		\epsilon_1(t-1)\\
		\epsilon_2(t-1)\\
		\epsilon_3(t-1)
	\end{bmatrix} + \Phimat_2\begin{bmatrix}
		\epsilon_1(t-2)\\
		\epsilon_2(t-2)\\
		\epsilon_3(t-2)
	\end{bmatrix} + \begin{bmatrix}
		e_1(t)\\
		e_2(t)\\
		e_3(t)
	\end{bmatrix},
\end{align*}
where $\Phi_1$ and $\Phi_2$ are the coefficient matrices. The independent error term $e_1(t)$, $e_2(t)$, and $e_3(t)$ follow multivariate normal distribution with mean vector $\zerovec$ and covariance matrix, $\Sigmat_e$.

The parameter estimation can be achieved by two separate steps as described in \shortciteN{Hong2015}. The first step is to use optimization algorithms to obtain maximum likelihood estimation for the mean parameters, $\muvec = (\mu_1, \mu_2, \mu_3)\tran, \kappavec=(\kappa_1, \kappa_2, \kappa_3)\tran, \etavec=(\eta_1, \eta_2, \eta_3)\tran, \upsilonvec=(\upsilon_1, \upsilon_2, \upsilon_3)\tran$, and $\zetavec=(\zeta_1, \zeta_2)\tran$. With the maximum likelihood estimation, the second step is to use multivariate least squares method to estimate the residuals and autocorrelations. For convenience, we denote the parameters for modeling covariate process as $\thetavec_{X}$, and its estimate is $\hat{\thetavec}_{X}$. Furthermore, the confidence interval of $\thetavec_X$ can be achieved by bootstrapping.

\section{Covariate Processes and Effect Functions For Simulation}\label{sec:cov.proc.eff.sim}
\begin{figure}
	\centering
	\begin{subfigure}{0.40\textwidth}
		\centering
		\includegraphics[width=\linewidth]{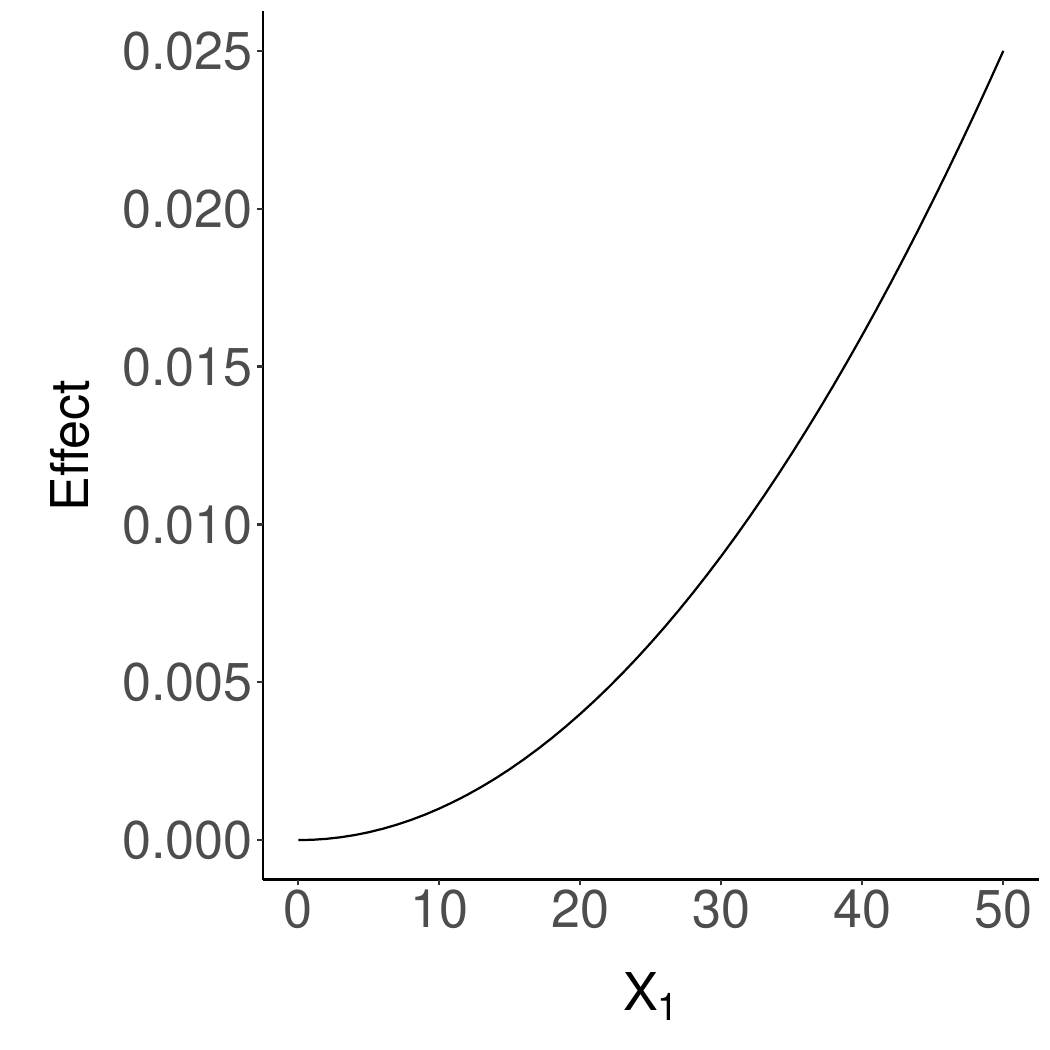}
	\end{subfigure}%
	\begin{subfigure}{0.40\textwidth}
		\centering
		\includegraphics[width=\linewidth]{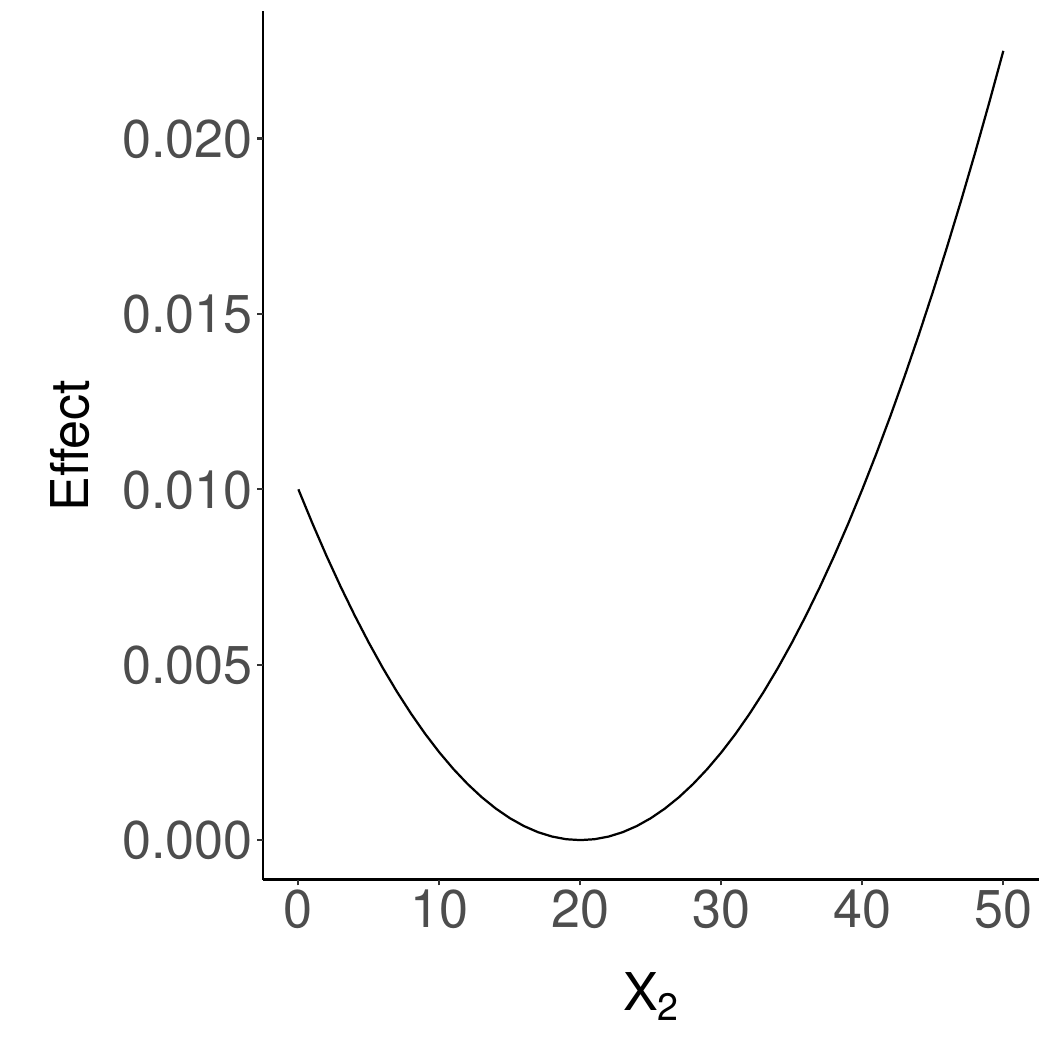}
	\end{subfigure}%
	\caption{The covariate effect functions for dynamic covariates $\Xmat_1$ and $\Xmat_2$.}
	\label{fig:supp.plot.sim.cov.eff}
\end{figure}

The simulated covariate processes, denoted as $\Xmat_1(t)$ and $\Xmat_2(t)$, are both functions of time $t$. The detailed forms of the covariate processes, adapted from \shortciteN{Xu2016}, are as follows,
\begin{align*}
	\Xmat_1(t) &= \mu_1 + \kappa_1 \sin \left[ \frac{2 \pi}{365}(t-\eta_1)\right] + \left\{1+v_1 \sin\left[\frac{2\pi}{365}(t-\zeta_1)\right] \right\}\epsilon_1(t),\\
	\Xmat_2(t) &= \mu_2 + \kappa_2 \cos \left[ \frac{2 \pi}{365}(t-\eta_2)\right] + \left\{1+v_2 \sin\left[\frac{2\pi}{365}(t-\zeta_2)\right] \right\}\epsilon_2(t),
\end{align*}
where $\epsilon_1(t)$ and $\epsilon_2(t)$ are i.i.d error terms that follow $\normal(0,4)$. Fixed parameters $(\mu_1, \kappa_1, \eta_1, v_1, \zeta_1)\\
 = (25,16,103,0.31,33)$ and $(\mu_2, \kappa_2, \eta_2, v_2, \zeta_2) = (25,18,103,0.31,33)$. We choose the effect functions of $\Xmat_1$ and $\Xmat_2$ as follows,
\begin{align*}
	\ffun_{1}(x) = \frac{1}{10^3} x^2 + 0.01, \text{ } \ffun_{2}(x) = \frac{1}{15 \times 10^3} (x-30)^2 + 0.01.
\end{align*}
Visualizations of $f_{1}(x)$ and $f_{2}(x)$ are presented in Figure~\ref{fig:supp.plot.sim.cov.eff}. Within the range of $X_1$ and $X_2$, $f_1(x)$ is increasing and $f_2(x)$ is convex.

\begin{table}
	\centering
	\caption{Summary of posterior medians for a subset of model parameters under different scenarios. The ``Median'' column represents the mean of posterior medians obtained from simulated datasets.}\label{tab:supp.sim.par.inf2}
	\begin{tabular}{ccccccc}\toprule
		$I$ & $n_i$ & Parameter & Median & Bias$\times10^{2}$ & SD$\times10^{2}$ & CP\\
		\midrule
		10 &  10 &  $\gamma_2$ & 1.119 & 5.359 & 10.540 & 0.925 \\
		&  20 &            & 1.103 & 2.893 & 3.446 & 0.929 \\
		&  50 &            & 1.102 & 1.553 & 1.944 & 0.951 \\
		\midrule
		20 &  10 & $\gamma_2$ & 1.099 & 3.044 & 3.988 & 0.955 \\
		&  20 &            & 1.104 & 1.906 & 2.559 & 0.952 \\
		&  50 &            & 1.101 & 1.059 & 1.321 & 0.961 \\
		\midrule
		50 &  10 &  $\gamma_2$ & 1.100 & 2.016 & 2.570 & 0.941 \\
		&  20 &            & 1.101 & 1.177 & 1.464 & 0.944 \\
		&  50 &            & 1.100 & 0.653 & 0.830 & 0.961 \\
		\midrule
		10 &  10 &  $\sigma_{2}$ & 0.101 & 2.368 & 2.887 & 0.906\\
		&  20 &         		 & 0.094 & 1.850 & 2.677 & 0.924 \\
		&  50 &          		 & 0.094 & 1.969 & 2.415 & 0.927 \\
		20 &  10 & $\sigma_{2}$ & 0.099 & 1.375 & 1.788 & 0.950 \\
		&  20 &         		& 0.098 & 1.206 & 1.657 & 0.949 \\
		&  50 &         		& 0.097 & 1.297 & 1.640 & 0.949 \\
		50 &  10 & $\sigma_{2}$ & 0.099 & 0.865 & 1.061 & 0.939 \\
		&  20 &          		& 0.100 & 0.789 & 1.022 & 0.945 \\
		&  50 &          		& 0.100 & 0.800 & 0.829 & 0.949 \\
		\bottomrule
	\end{tabular}
\end{table}

\section{Additional Simulation Results}

\begin{figure}
	\centering
	\includegraphics[width=1\linewidth]{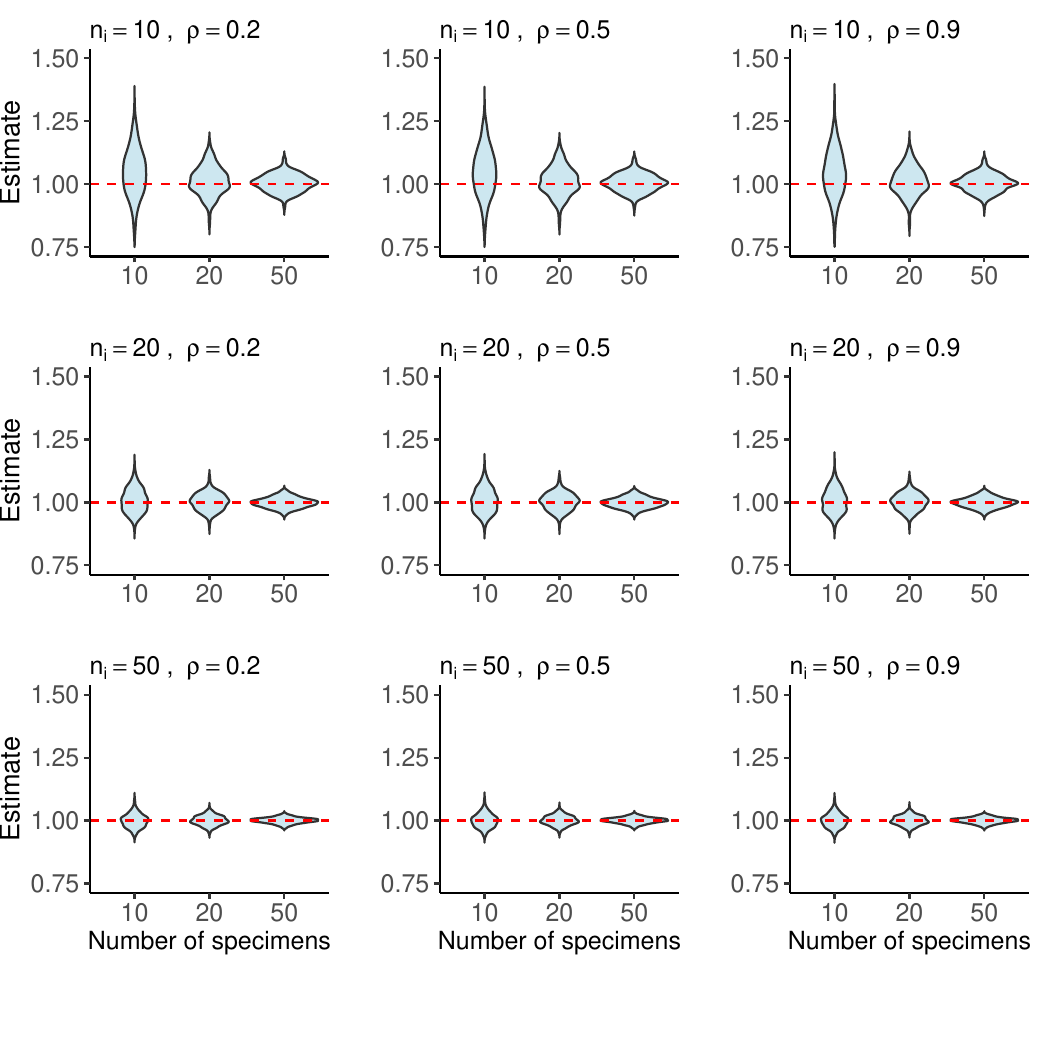}
	\caption{Violin plot for $\gamma_1$ given different scenarios.}
	\label{fig:supp.sim.r1.plt}
\end{figure}

\begin{figure}
	\centering
	\includegraphics[width=1\linewidth]{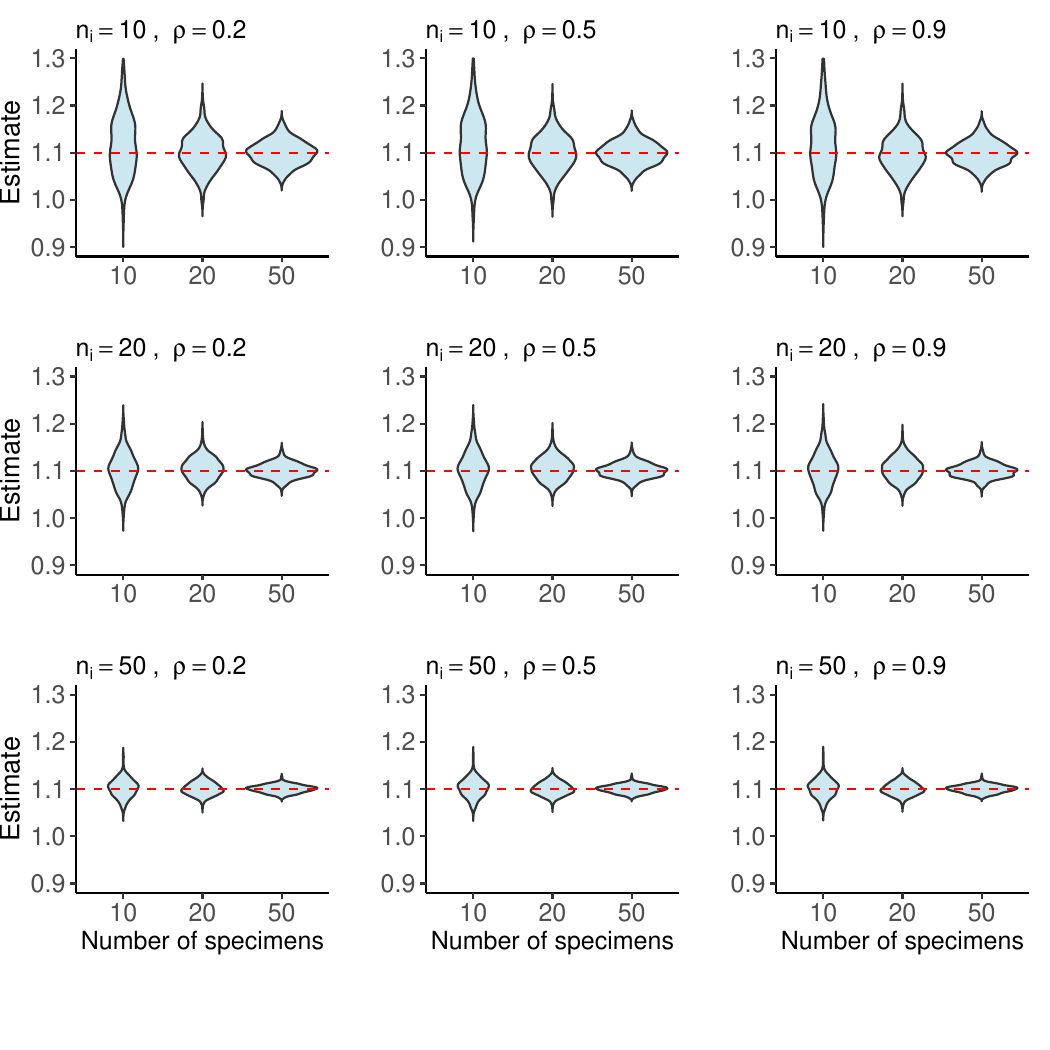}
	\caption{Violin plot for $\gamma_2$ given different scenarios.}
	\label{fig:supp.sim.r2.plt}
\end{figure}

\begin{figure}
	\centering
	\includegraphics[width=1\linewidth]{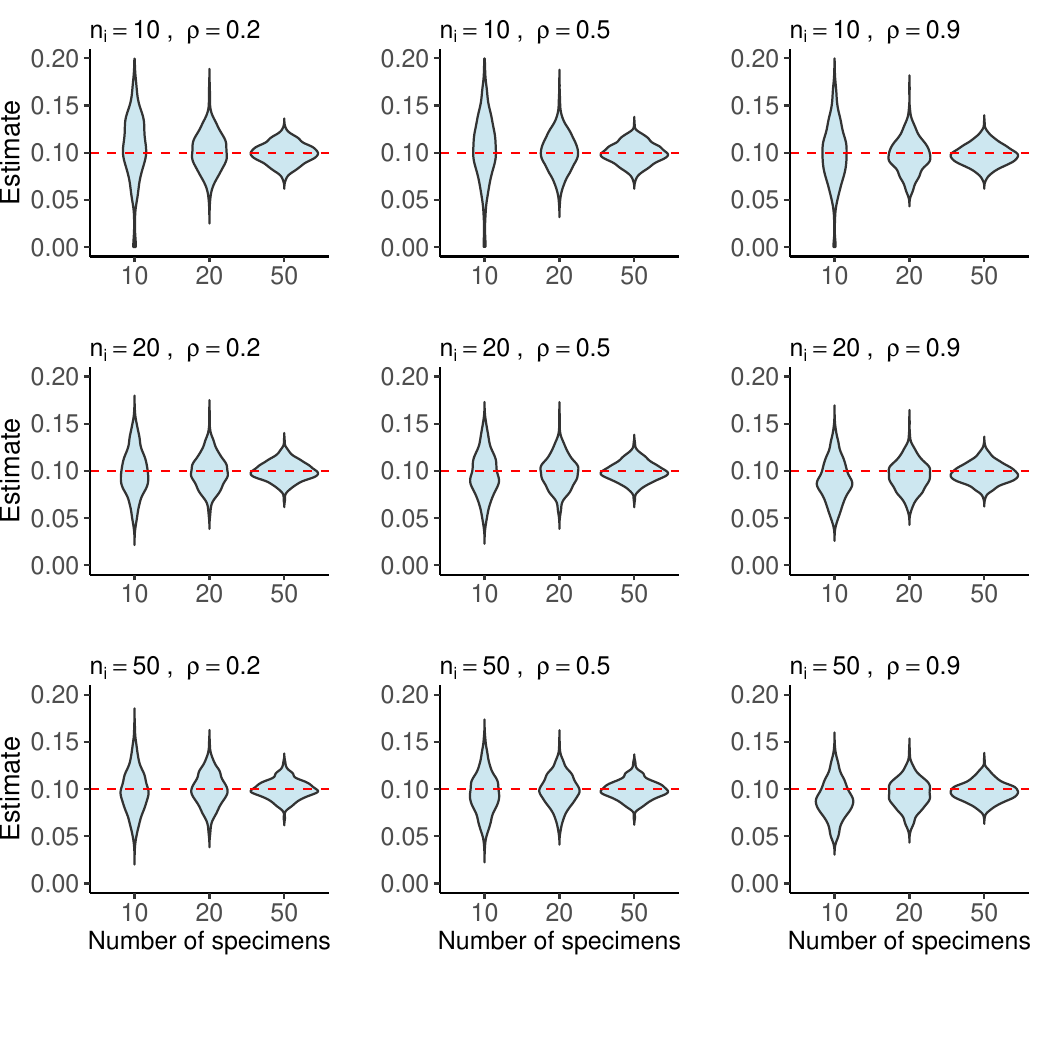}
	\caption{Violin plot for $\sigma_1$ given different scenarios.}
	\label{fig:supp.sim.sig_w1.plt}
\end{figure}

\begin{figure}
	\centering
	\includegraphics[width=1\linewidth]{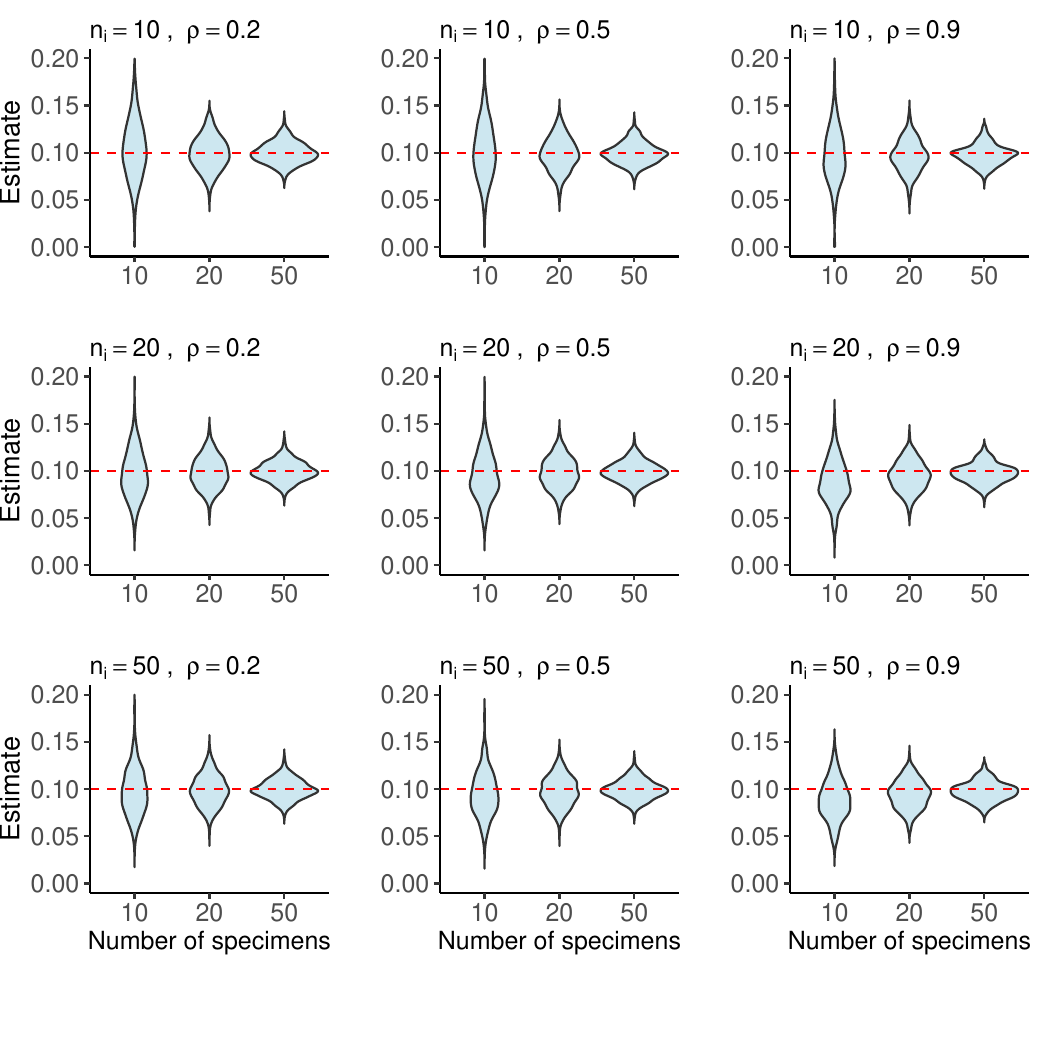}
	\caption{Violin plot for $\sigma_2$ given different scenarios.}
	\label{fig:supp.sim.sig_w2.plt}
\end{figure}

Table~\ref{tab:supp.sim.par.inf2} shows the summary of posterior medians for a subset of model parameters under different scenarios, in which the ``Median'' column represents the mean of posterior medians obtained from simulated datasets. Figures~\ref{fig:supp.sim.r1.plt},~\ref{fig:supp.sim.r2.plt},~\ref{fig:supp.sim.sig_w1.plt}, and~\ref{fig:supp.sim.sig_w2.plt} illustrate the estimated distributions for parameter estimators of $\gamma_1$, $\gamma_2$, $\sigma_1$, and $\sigma_{2}$ given different combinations of $I$, $n_i$, and $\rho$. Estimation improves with an increase in $I$, as shown by the violin plots, where higher $I$ values show greater concentration around the true value.

\begin{figure}
	\centering
	\includegraphics[width=1\linewidth]{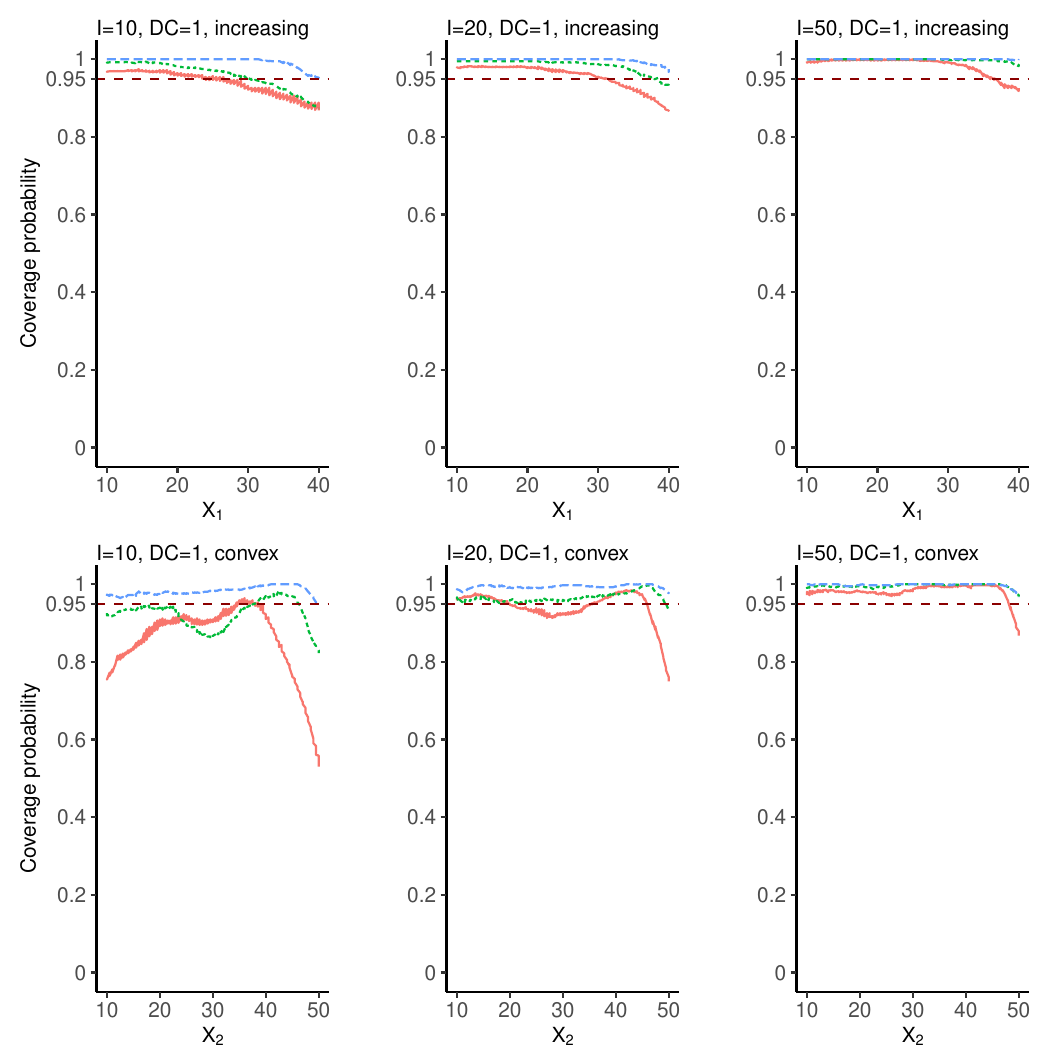}
	\caption{CP plots for effect functions of DC 1. On the $y$-axis, each point represents the CP for the effect of a given covariate value.}
	\label{fig:supp.sim.effect_cp1.plt}
\end{figure}

\begin{figure}
	\centering
	\includegraphics[width=1\linewidth]{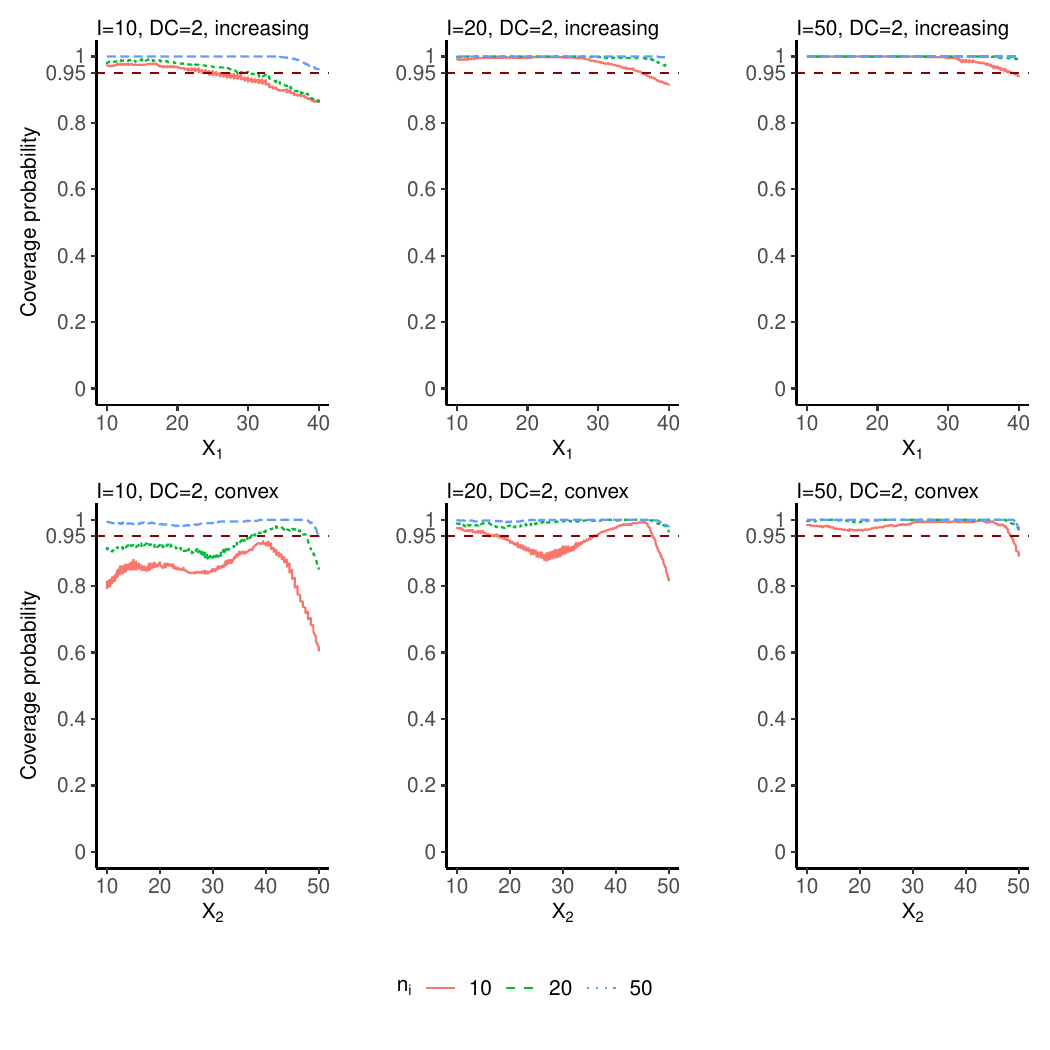}
	\caption{CP plots for effect functions of DC 2. On the $y$-axis, each point represents the CP for the effect of a given covariate value.}
	\label{fig:supp.sim.effect_cp2.plt}
\end{figure}

Figures~\ref{fig:supp.sim.effect_cp1.plt} and~\ref{fig:supp.sim.effect_cp2.plt} show the CP of effect functions across different shapes, DCs, and scenarios. For both shapes, improved CP can be achieved by increasing $I$ or $n_i$, as indicated by the lines moving closer to 1 with increases in $I$ or $n_i$. Due to an insufficient amount of extreme data points, the CPs fall below 95\% when covariate values are large. In all scenarios, the CPs for the increasing shape perform better than those of the convex shape. This is expected, as more complex shapes, like the convex form, are typically more challenging to estimate.

\begin{figure}
	\centering
	\includegraphics[width=1\linewidth]{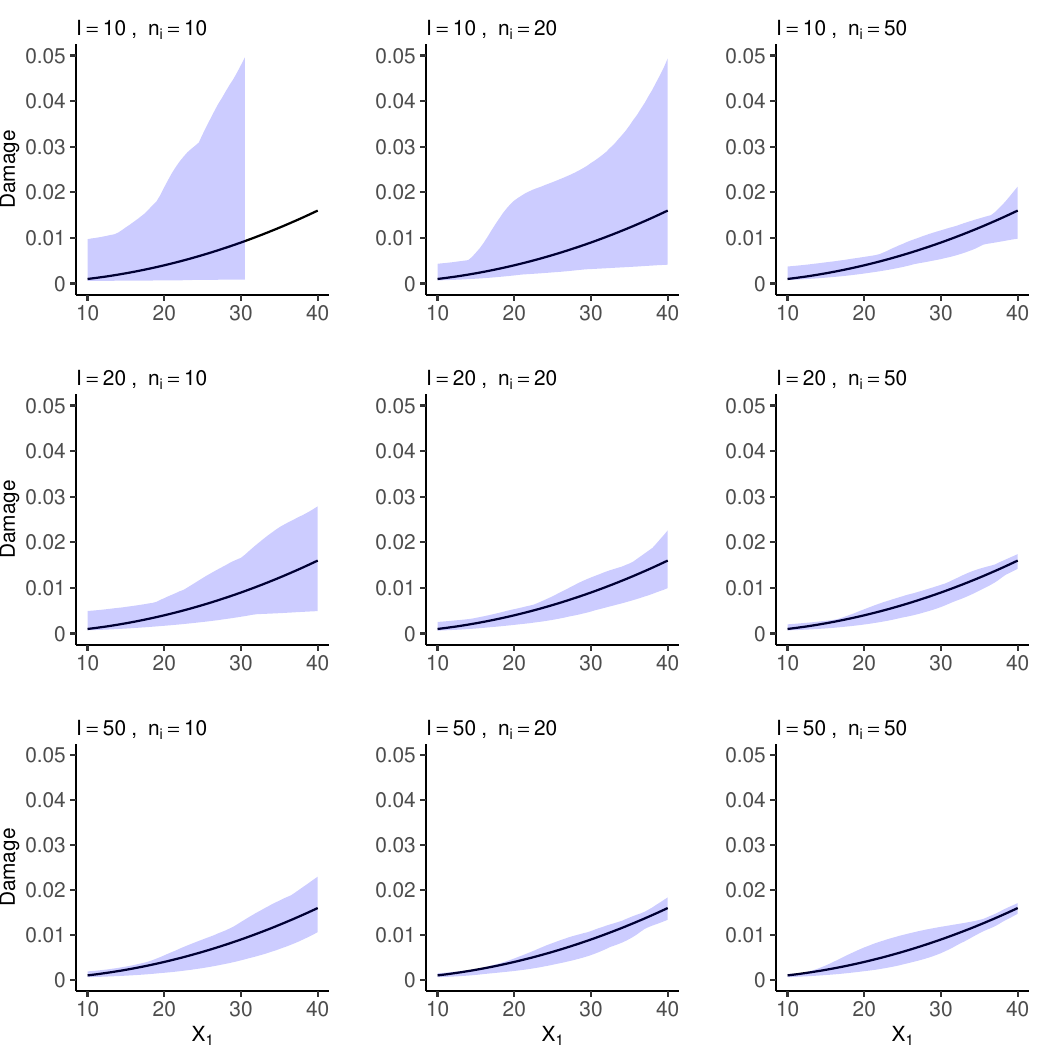}
	\caption{Effect function (increasing) coverage plots given different $I$ and $n_i$ for DC 1. The blue covered area is constructed by the estimated effect function from different simulated datasets. The black solid line represents true effect function.
	}
	\label{fig:supp.sim.effect_inc_1_coverage.plt}
\end{figure}

\begin{figure}
	\centering
	\includegraphics[width=1\linewidth]{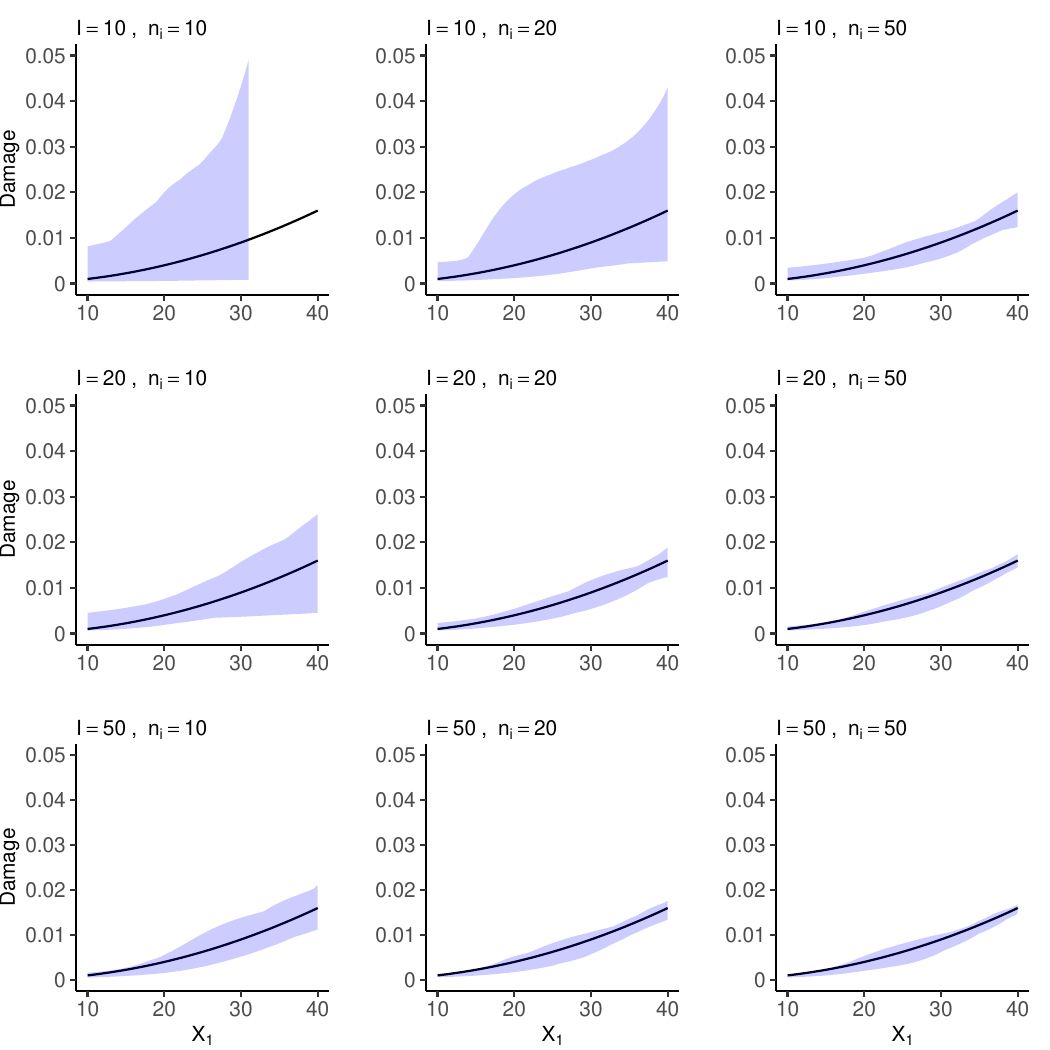}
	\caption{Effect function (increasing) coverage plots given different $I$ and $n_i$ for DC 2. The blue covered area is constructed by the estimated effect functions from different simulated datasets. The black solid line represents true effect function.
	}
	\label{fig:supp.sim.effect_inc_2_coverage.plt}
\end{figure}

\begin{figure}
	\centering
	\includegraphics[width=1\linewidth]{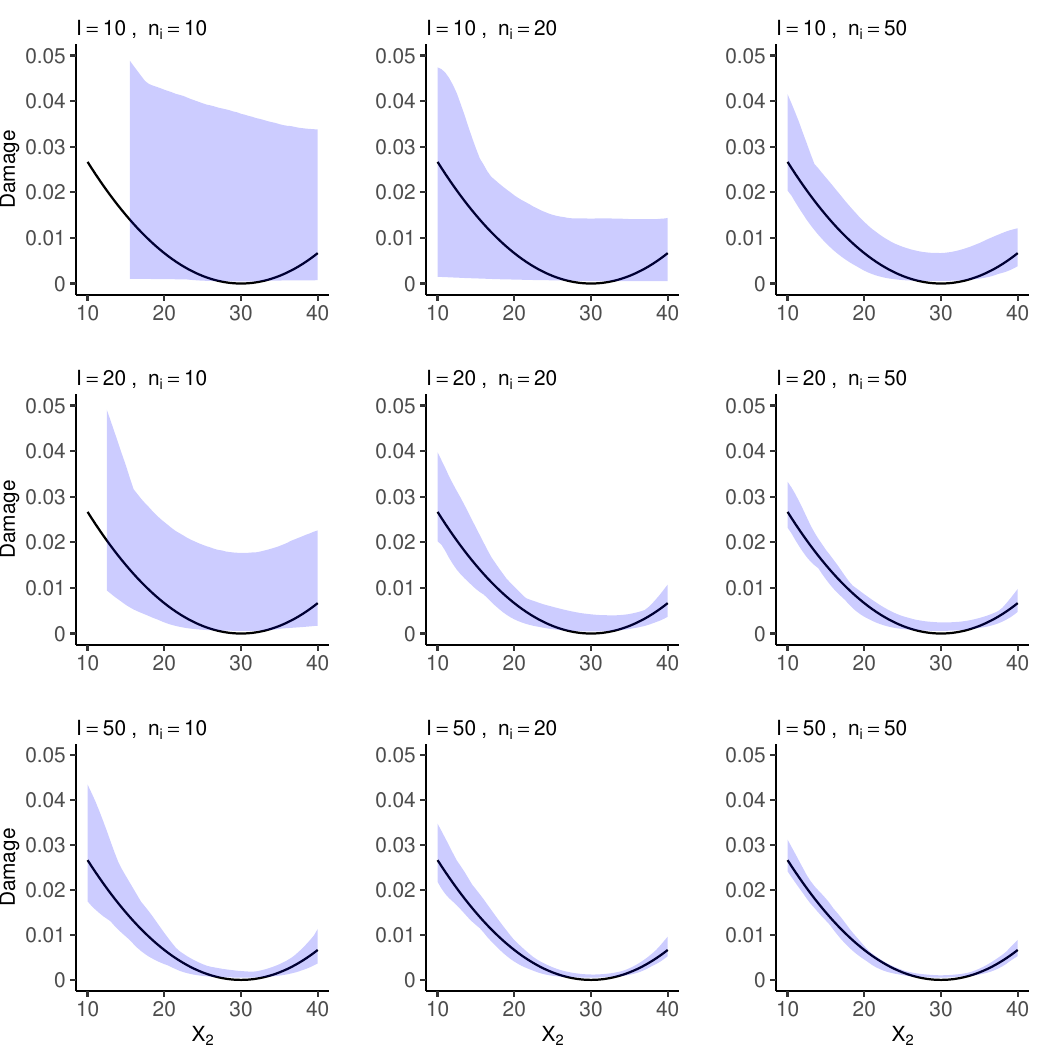}
	\caption{Effect function (convex) coverage plots given different $I$ and $n_i$ for DC 1. The blue covered area is constructed by the estimated effect functions from different simulated datasets. The black solid line represents true effect function.
	}
	\label{fig:supp.sim.effect_convex_1_coverage.plt}
\end{figure}

\begin{figure}
	\centering
	\includegraphics[width=1\linewidth]{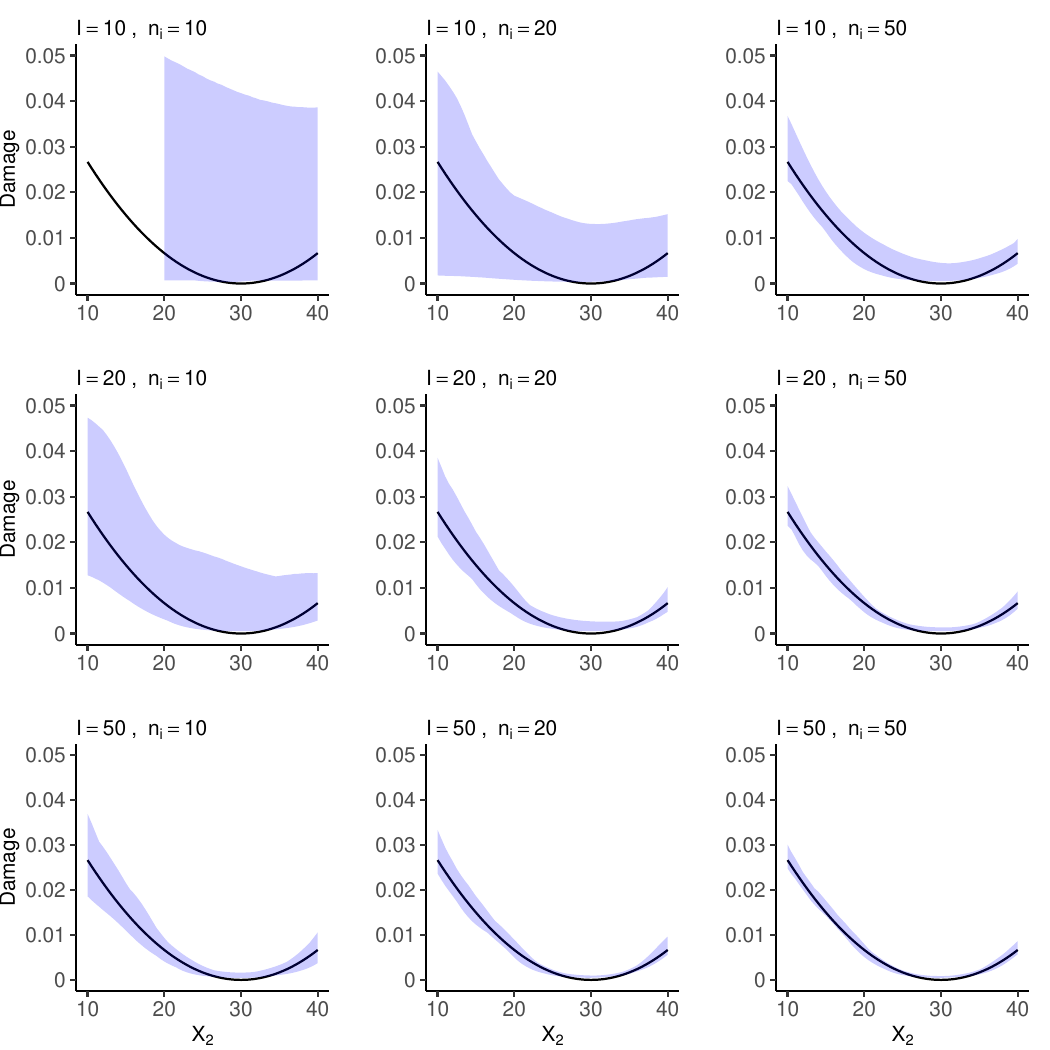}
	\caption{Effect function (convex) coverage plots given different $I$ and $n_i$ for DC 2. The blue covered area is constructed by the estimated effect functions from different simulated datasets. The black solid line represents true effect function.
	}
	\label{fig:supp.sim.effect_convex_2_coverage.plt}
\end{figure}

Additionally, Figures~\ref{fig:supp.sim.effect_inc_1_coverage.plt},~\ref{fig:supp.sim.effect_inc_2_coverage.plt},~\ref{fig:supp.sim.effect_convex_1_coverage.plt}, and~\ref{fig:supp.sim.effect_convex_2_coverage.plt} present ribbon plots for effect function estimation. These plots validate conclusions from the above discussion; the ribbons concentrate and align more closely with the true effect function as $I$ or $n_i$ increase. The right tails of the ribbons are wider, indicating greater uncertainty at extreme covariate values, and ribbons for the increasing shape cover the true effect function more accurately compared to those from the convex shape.

\begin{figure}
	\centering
	\includegraphics[width=1\linewidth]{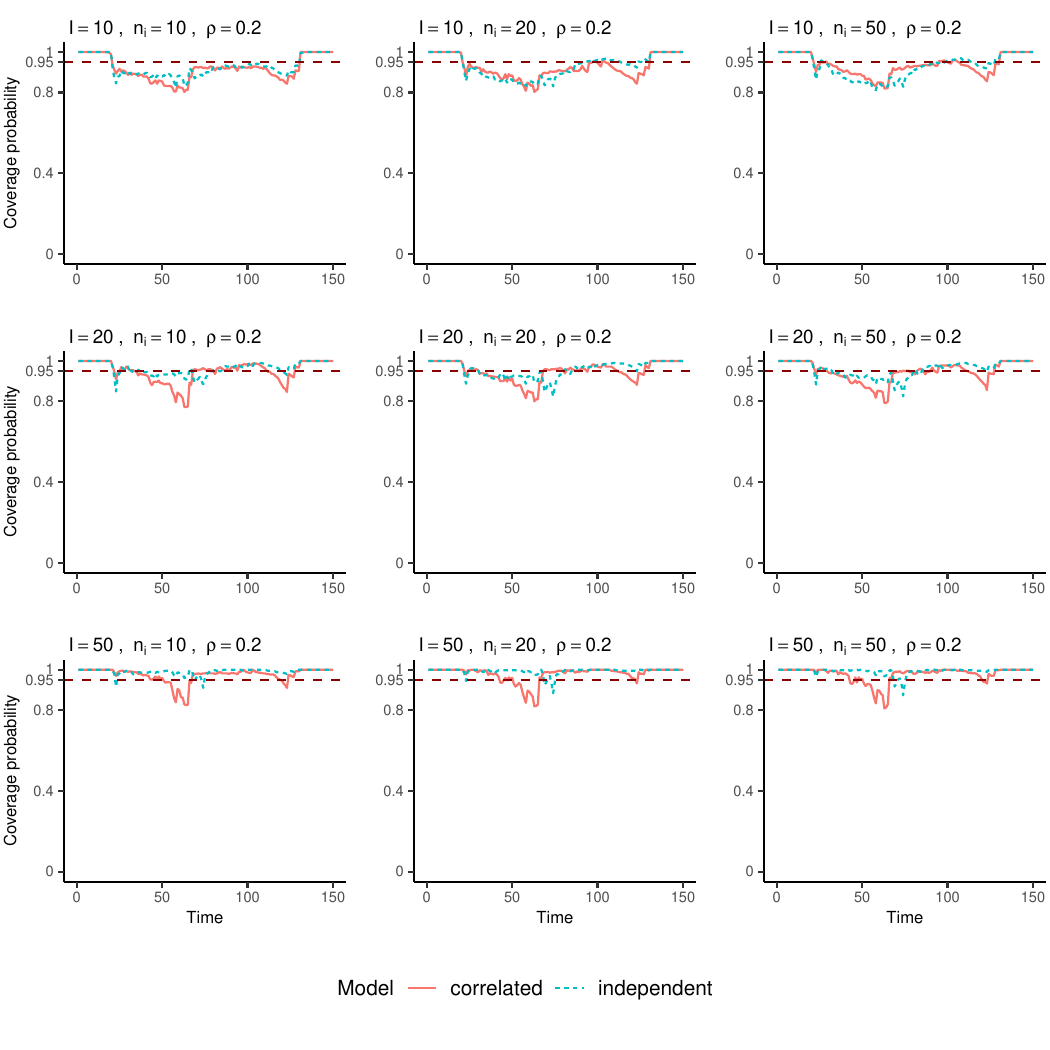}
	\caption{CP plots for failure time distribution given $\rho=0.2$. On the $y$-axis, each point represents the CP for the failure probability of a given time.}
	\label{fig:supp.sim.dft_cp1.plt}
\end{figure}

\begin{figure}
	\centering
	\includegraphics[width=1\linewidth]{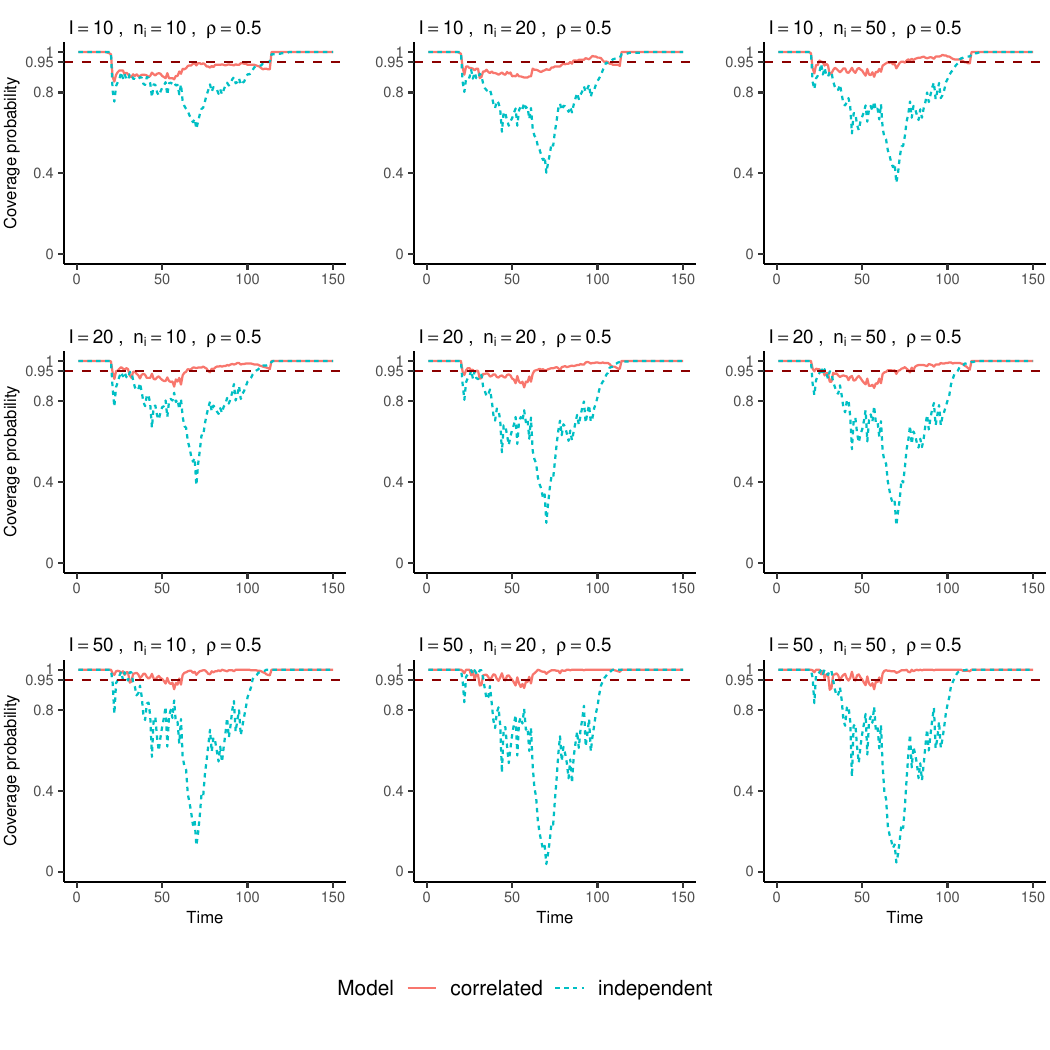}
	\caption{CP plots for failure time distribution given $\rho=0.5$. On the $y$-axis, each point represents the CP for the failure probability of a given time.}
	\label{fig:supp.sim.dft_cp2.plt}
\end{figure}

\begin{figure}
	\centering
	\includegraphics[width=1\linewidth]{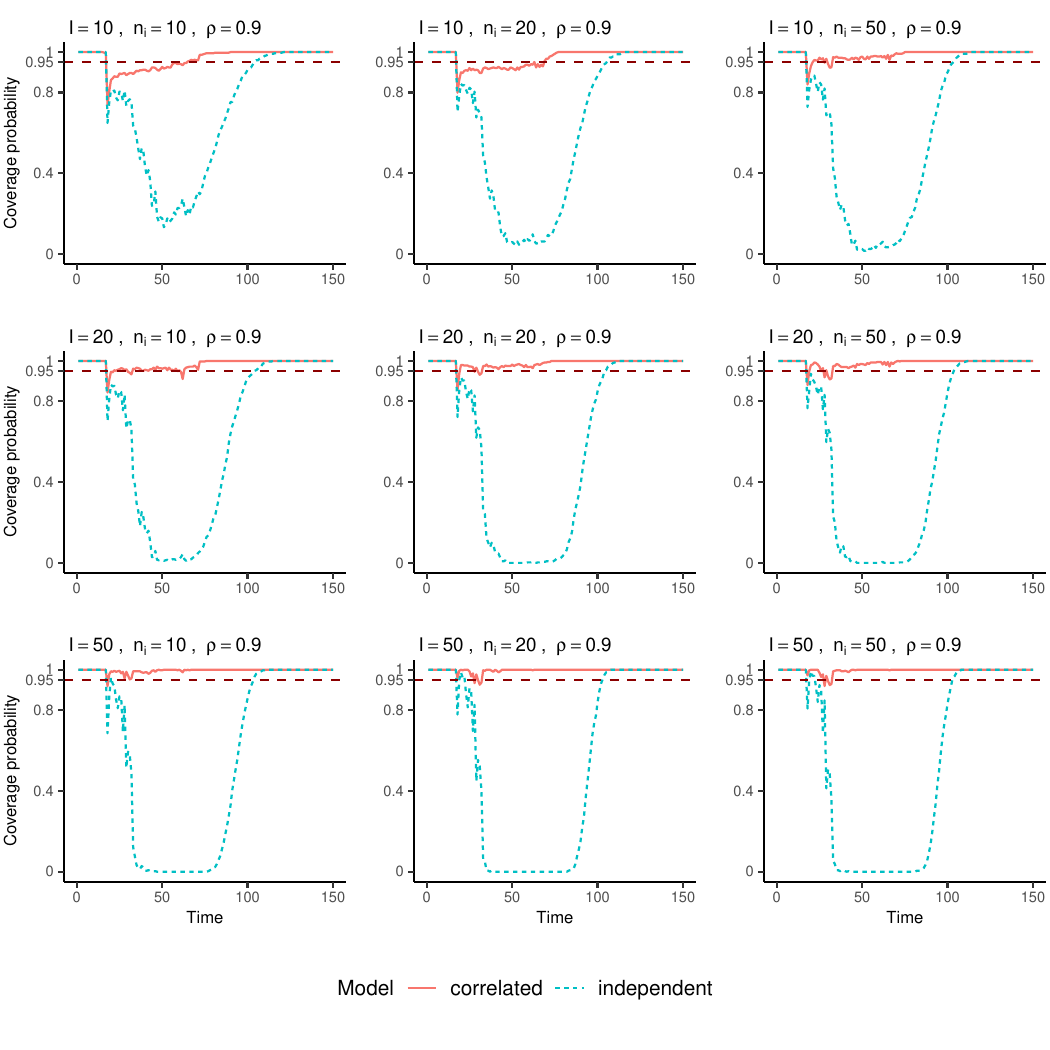}
	\caption{CP plots for failure time distribution given $\rho=0.9$. On the $y$-axis, each point represents the CP for the failure probability of a given time.}
	\label{fig:supp.sim.dft_cp3.plt}
\end{figure}

Figures~\ref{fig:supp.sim.dft_cp1.plt},~\ref{fig:supp.sim.dft_cp2.plt}, and~\ref{fig:supp.sim.dft_cp3.plt} present CP plots regarding to failure time distribution estimation. Under conditions of weak correlation, the independent and the correlated models exhibit similar performance, with their CPs closely aligned and approximating 95\%. However, under the conditions of medium to strong correlation, the correlated model significantly outperforms the independent model. In these scenarios, the CPs of the correlated model maintain close or higher than 95\%, while the CPs of the independent model fall drastically below 80\% and approach 0\% under strong correlations.



\end{document}